\documentclass{PoS}

\def\be{\begin{equation}}
\def\ee{\end{equation}}
\def\bea{\begin{eqnarray}}
\def\eea{\end{eqnarray}}

\title{Topics in Cosmology}

\ShortTitle{Topics in Cosmology}

\author{\speaker{Robert Brandenberger}\\
        Physics Department, McGill University, Montreal, QC, H3W 2T8, Canada\\
        E-mail: \email{rhb@hep.physics.mcgill.ca}}

\abstract{These lectures present a brief review of inflationary cosmology,
provide an overview of the theory of cosmological perturbations, and then
focus on the conceptual problems of the current paradigm of early universe
cosmology, thus motivating an exploration of the potential of string theory
to provide a new paradigm. Specifically, the string gas cosmology model
is introduced, and a resulting mechanism for structure formation which
does not require a period of cosmological inflation is discussed.}

\FullConference{School on Particle Physics, Gravity and Cosmology\\
		21 August - 2 September 2006 \\
		Dubrovnik, Croatia}

\begin{document}

\section{Introduction and Overview}

These lectures focus on three topics. The first is an overview
of the current paradigm of early universe cosmology, the
{\it inflationary universe scenario}. The second topic is
a pedagogical presentation of the theory of cosmological perturbations,
the main tool of modern cosmology which allows us to connect
theories of the very early universe with observational data.
The third topic is {\it string gas cosmology}, an attempt to
construct a new scenario of the very early universe based on
established principles of string theory.

Over the past two and a half decades, cosmology has become a
science dominated by data of rapidly increasing accuracy. Today,
we have three-dimensional maps of the distribution of galaxies in 
space which contain more than one hundred thousand galaxies 
\cite{2dF,SDSS}. They clearly indicate that luminous matter in
the universe is neither uniformly nor randomly distributed. There
are clear patterns to be seen: clusters of galaxies, superclusters,
filaments and voids (regions of space empty of galaxies). The
distribution can be quantified in terms of the luminosity power
spectrum. A key challenge for cosmology is to understand the
origin of these patterns in the distribution of matter.

Another observational window in cosmology is the cosmic microwave
background (CMB) radiation. Overall, this radiation is characterized
by a surprising isotropy. At a fractional level of a bit less
than $10^{-4}$, however, there are anisotropies. These
can be quantified in terms of their angular power spectrum. First, the
sky map (two-dimensional) of anisotropies is expanded in spherical harmonics
$Y_{lm}$:
\be
{{\Delta T} \over T}(\theta, \varphi) \, = \, 
\sum_{l=1}^{\infty} \sum_{m = -l}^{l} a_{lm} Y_{lm}(\theta, \varphi)
\ee
where $\theta$ and $\phi$ are the usual angles on the surface of the sphere.
If the fluctuations are due to a Gaussian random process with no
distinguished direction in the sky, then the complete information about
the fluctuations is given by the ensemble average (denoted by pointed
parentheses) of the coefficients $a_{lm}$:
\be
c_l \, = \, <|a_{lm}|^2> \, ,
\ee
These $c_l$ coefficients define the angular power spectrum of CMB
anisotropies. Figure 1 is the full sky map of CMB anisotropies from the
WMAP experiment \cite{WMAP}. Figure 2 shows the resulting angular power
spectrum )(what is plotted on the vertical axis is $l(l + 1)C_l / (2 \pi)$). 
The key features are the flat region at small values of $l$ (large
angular scales) and the characteristic oscillations of the spectrum at
intermediate angular scales. Another crucial challenge for cosmology is to
explain both the overall isotropy of the CMB, and the specific patterns
of anisotropies.

\begin{figure}
\centering
%\centerline{\epsfxsize=3in\epsfbox{fig4.ps}}
\includegraphics[height=8cm]{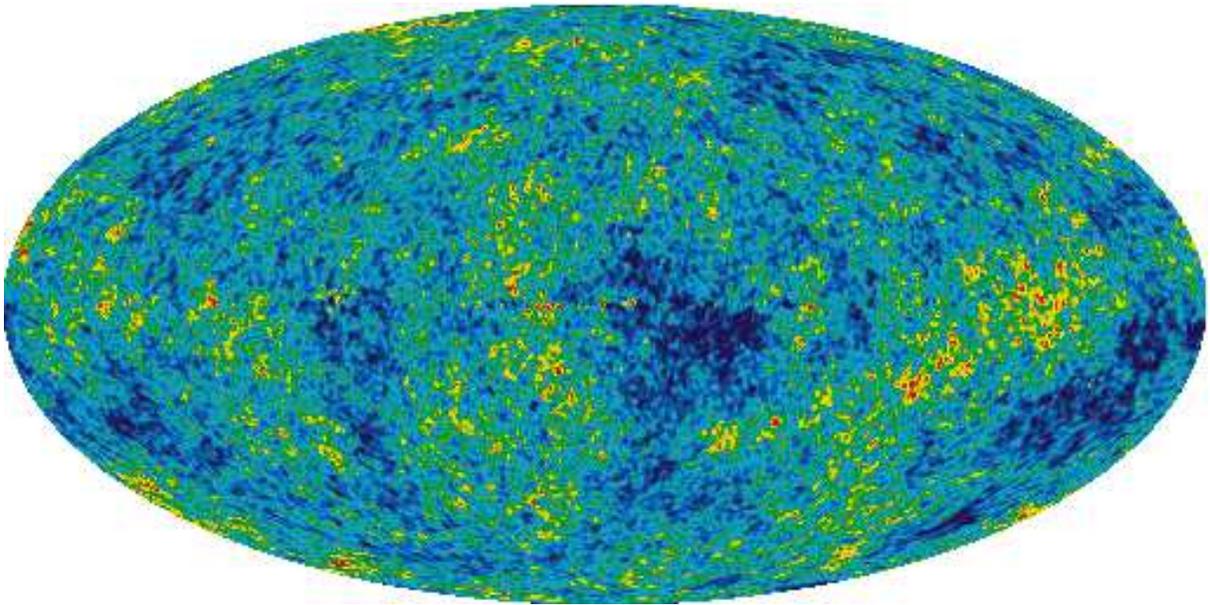}
\caption{All sky map of the temperature anisotropies in the CMB from the
WMAP satellite experiment. Credit: NASA/WMAP Science Team.}
\label{fig:1}       
\end{figure}

\begin{figure}
\centering
%\centerline{\epsfxsize=3in\epsfbox{canc1.eps}}
\includegraphics[height=6cm]{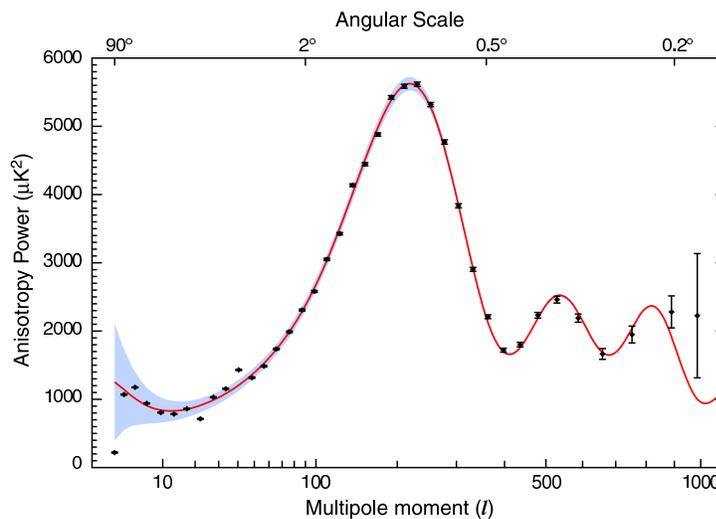}
\caption{The power spectrum of CMB anisotropies as computed from the
WMAP satellite experiment. The horizontal axis is the angular
quantum number $l$, the vertical
axis gives the power of the CMB on the respective scales.
The dots represent the data points (with their errors bars
indicated), the solid curve is the prediction of the best-fit inflationary
model. Credit: NASA/WMAP Science Team.}
\label{fig:2}       
\end{figure}

According to our present understanding, we must look to the very early 
universe to find an explanation for the observed structures. The reason
is that in Standard Big Bang cosmology (SBB), which well describes the
cosmological evolution at late times (times later and including the period
of nucleosynthesis) the physical wavelength of fixed comoving scales is
increasing less fast than the Hubble radius (an important length scale
which is defined and whose role is described at the end of this section). 
Hence, the scales which are
currently observed were outside of the Hubble radius at early times,
and no causal structure formation scenario is possible \footnote{Topological
defect models \cite{ShellVil,HK,RHBrev0} provide a way to circumvent
this reasoning. They, however, also involve new physics of the very early
universe.}.

It is a remarkable success of inflationary cosmology, our current paradigm of
early universe cosmology, that it, in addition to explaining why the
universe is large, spatially flat, and containing a large amount of entropy,
provides a causal mechanism for the origin of inhomogeneities in the
universe. The solid curve in Figure 2 represents the predictions of
inflationary cosmology \footnote{Several cosmological parameters, e.g. the
current value of the cosmological constant and the fractional baryon density,
have to be fixed in order to obtain this excellent agreement. The number of
free parameters, however, is much, much smaller than the number of data
points.} The predictions were made more than 15 years before the data shown
in Figure 1 were collected.

In these lectures, however, I would like to focus less on the phenomenological
successes of the inflationary paradigm, but more on the conceptual problems
which current realizations of inflationary cosmology are confronted with.
These problems call out for the development of a new paradigm, a paradigm
which must include new fundamental physics, a theory which better
describes space, time and matter and the highest densities. 
The best current candidate for such a theory is superstring theory. Hence,
in the third main part of these lectures (Section 4) I will explore the 
possibility that string theory may lead to a new paradigm of early universe 
cosmology. My approach to string cosmology in this section is complementary
to the one taken in the lectures of Cliff Burgess at this school (see
\cite{Cliff}), in which avenues of obtaining inflation
in the context of models coming from string theory are explored (see also
\cite{Jim,Andrei} for other reviews of such avenues).

The theory of cosmological perturbations plays a key role in modern cosmology,
since it provides the techniques with which to calculate, in the context
of any given scenario of the very early universe, the generation and
evolution of the predicted inhomogeneities in the matter distribution and
anisotropies in the CMB, and thus allows for a comparison between fundamental
theory and observational data. The second part of these lectures (Section 3)
provides a pedagogical overview of this theory.

I begin, however, with a discussion of inflationary cosmology, the
current paradigm of cosmology.

To establish our notation and framework, we
will be taking the background space-time to be homogeneous
and isotropic, with a metric given by
\begin{equation} \label{background}
ds^2 \, = \, dt^2 - a(t)^2 d{\bf x}^2 \, ,
\end{equation}
where $t$ is physical time, $d{\bf x}^2$ is the Euclidean metric
of the spatial hypersurfaces (here taken for simplicity to be
spatially flat), and $a(t)$ is the scale factor. The scale
factor determines the {\it Hubble expansion rate} via 
\begin{equation}
H(t) \, = \, {{\dot{a}} \over a}(t) \, .
\end{equation}
The coordinates ${\bf x}$ used above are {\it comoving} coordinates,
coordinates painted onto the expanding spatial hypersurfaces.

We are interested in tracking the time evolution of the physical
wavelength of the currently observed patterns in the distribution
of matter and of the CMB anisotropies. Since the patterns are
assumed to be frozen in in comoving coordinates, the physical
wavelength scales as $a(t)$. 

A key length scale in cosmology is the {\it Hubble radius}
\begin{equation}
l_H(t) \, = \, H^{-1}(t) \, ,
\end{equation}
defined to be the inverse Hubble expansion rate. As will be explained later, 
the Hubble radius is the maximal distance that 
microphysics can act coherently over a Hubble expansion time -
in particular it is the maximal distance on which any causal process could
create fluctuations.

\section{Inflationary Cosmology: The Current Paradigm}

Standard Big Bang (SBB) cosmology, the precursor to inflationary cosmology
as the paradigm for the evolution of the early universe, is based on
a classical physics description of both space-time (via
Einstein's theory of General Relativity) and matter (a superposition
of two perfect fluids, the first describing pressureless matter - cold
matter - the second describing radiation - the CMB). 
The key phenomenological success of the SBB model is the prediction of
the existence and black body nature of the CMB, the black body nature
of which was confirmed with spectacular accuracy by the COBE satellite
\cite{COBE} and UBC rocket experiments \cite{Halpern}.

On the other hand, the SBB scenario leaves many crucial questions
un-answered \cite{Guth}. Why is the universe so close to spatially flat?
Why is it so large and contains such a large entropy? Why is the
CMB isotropic to an accuracy of better than $10^{-4}$ (after
subtracting the dipole contributions due to our motion relative to the
rest frame of the CMB and the effects due to the emission of our own galaxy)?
Most importantly, what is the origin of the observed inhomogeneities in
the distribution of matter and of the small CMB anisotropies? These are
the ``flatness", ``entropy", ``horizon" and ``structure formation"
problems of SBB cosmology.
SBB also suffers from conceptual problems: the initial cosmological
singularity tells us both that the theory must be incomplete, and that
it is based on using the wrong fundamental physics input close to the
singularity \footnote{As we will see later on, these conceptual problems
persist in inflationary cosmology.}.

Inflationary cosmology \cite{Guth} (see also \cite{Sato,Brout,Starob}) 
provides an solution of the horizon, flatness and entropy
problems. In addition, it provides a mechanism for the origin of
structure in the universe based on causal physics \cite{ChibMukh}
(see also \cite{Press,Sato}). 

The idea of inflationary cosmology is to assume that there was
a period in the very early Universe during which the scale factor was
accelerating, i.e. ${\ddot a} > 0$. This implies that the Hubble
radius was shrinking in comoving coordinates, or, equivalently, that fixed
comoving scales were ``exiting'' the Hubble radius. In the simplest models of
inflation, the scale factor increases nearly exponentially.
\begin{figure}
\centering
%\centerline{\epsfxsize=3in\epsfbox{canc1.eps}}
\includegraphics[height=6cm]{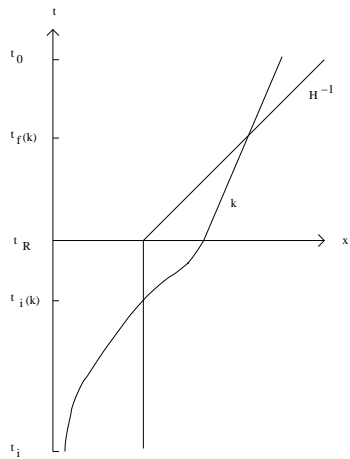}
\caption{Space-time diagram (sketch) showing the evolution
of scales in inflationary cosmology. The vertical axis is
time, and the period of inflation lasts between $t_i$ and
$t_R$, and is followed by the radiation-dominated phase
of standard big bang cosmology. During exponential inflation,
the Hubble radius $H^{-1}$ is constant in physical spatial coordinates
(the horizontal axis), whereas it increases linearly in time
after $t_R$. The physical length corresponding to a fixed
comoving length scale labelled by its wavenumber $k$ increases
exponentially during inflation but increases less fast than
the Hubble radius (namely as $t^{1/2}$), after inflation.}
\label{fig:3}       
\end{figure}

As illustrated in Figure \ref{fig:3}, 
the basic geometry of inflationary cosmology
provides a solution of the fluctuation problem. As long as the phase
of inflation is sufficiently long, all length scales within our present
Hubble radius today originate at the beginning of inflation with a 
wavelength smaller than the Hubble radius at that time. Thus, it
is possible to create perturbations locally using physics obeying
the laws of special relativity (in particular causality). As will be
discussed later, it is quantum vacuum fluctuations of matter fields
and their associated curvature perturbations which are predicted
to be responsible for the structure we observe today.

Postulating a phase of inflation in the very early universe solves
the {\it horizon problem} of the SBB, namely it explains how the causal 
horizon at the time $t_{rec}$ when photons last scatter can be 
larger than the radius of the
past light cone at $t_{rec}$, the part of the last scattering surface which is 
visible today in CMB experiments. Inflation explains the near flatness of
the universe: in a decelerating universe spatial flatness is an unstable fixed
point of the dynamics, whereas in an accelerating universe it becomes an
attractor. Another important feature of inflation is that the volume of
space increases exponentially at constant energy density. If
this energy density is successfully converted to ordinary matter at the end
of the period of inflation, then the entropy of the universe is
exponentially larger after compared to before inflation. In addition,
with the exponential expansion of space it is easy to produce a universe of
our size today from a Planck scale universe at the initial Planck time,
something which is not possible in the SBB model. Thus inflation explains
the large current size and entropy of the universe.

Let us now consider how it is possible to obtain a phase of cosmological
inflation. We will assume that space-time is described using the
equations of General Relativity \footnote{Note, however, that the first model
of exponential expansion of space \cite{Starob} made use of a higher derivative
gravitational action.}. In this case, the dynamics of the scale factor
$a(t)$ is determined by the Friedmann-Robertson-Walker (FRW) equations
\begin{equation} \label{FRW1}
({{\dot a} \over a})^2 \, = \, 8 \pi G \rho 
\end{equation}
and
\begin{equation} \label{FRW2}
{{\ddot a} \over a} \, = \, - 4 \pi G (\rho + 3p)
\end{equation}
where for simplicity we have omitted the contributions of spatial curvature
(since spatial curvature is diluted during inflation) and of the cosmological
constant (since any small cosmological constant which might be present today
has no effect in the early Universe since the associated energy density does
not increase when going into the past). In the above, $\rho$ and $p$ denote
the energy density and pressure, respectively. From (\ref{FRW2}) it is clear
that in order to obtain an accelerating universe, matter with sufficiently
negative pressure
\begin{equation}
p \, < \, - {1 \over 3} \rho
\end{equation}
is required. Exponential inflation is obtained for $p = - \rho$.

Conventional perfect fluids have positive semi-definite pressure and thus
cannot yield inflation. However, we know that a description of matter in
terms of classical perfect fluids must break down at early times. An improved
description of matter will be given in terms of quantum fields. Scalar
matter fields are special in that they allow at the level of a renormalizable
action the presence of a potential energy term. The energy density and
pressure of a scalar field $\varphi$ with canonically normalized action
\footnote{See \cite{kinflation} for a discussion of fields with non-canonical
kinetic terms.}
\begin{equation} \label{scalarlag}
S \, = \, \int d^4x \sqrt{-g}
\bigl[{1 \over 2} \partial_{\mu} \varphi \partial^{\mu} \varphi
- V(\varphi)\bigr] \,  
\end{equation}
(where Greek indices label space-time coordinates, $V(\varphi)$ is the
potential energy density, and $g$ is the determinant of the metric) 
are given by
\begin{eqnarray}
\rho \, &=& \, {1 \over 2} ({\dot \varphi})^2 + 
{1 \over 2} a^{-2} (\nabla \varphi)^2 + V(\varphi) \nonumber \\
p \, &=& \, {1 \over 2} ({\dot \varphi})^2 - 
{1 \over 6} a^{-2} (\nabla \varphi)^2
- V(\varphi) \, .
\end{eqnarray}
Thus, it is possible to obtain an almost exponentially expanding universe 
provided the scalar field configuration \footnote{The scalar field
yielding inflation is called the {\it inflaton}.} satisfies
\begin{eqnarray}
{1 \over 2} (\nabla_p \varphi)^2 \, &\ll& \, V(\varphi) \, , \label{gradcond} \\
{1 \over 2} ({\dot \varphi})^2 \, &\ll& \, V(\varphi) \, . \label{tempcond}
\end{eqnarray}
In the above, $\nabla_p \equiv a^{-1} \nabla$ is the gradient with respect
to physical as opposed to comoving coordinates.
Since spatial gradients redshift as the universe expands, the first condition
will (for single scalar field models) always be satisfied if it is satisfied at
the initial time \footnote{In fact, careful studies \cite{Kung}
show that since the gradients decrease even in a non-inflationary 
backgrounds, they can become subdominant even if they initially 
dominate.}. It is the second condition which is harder to satisfy. In
particular, this condition is in general not preserved in time even it is
initially satisfied \cite{Goldwirth}.

It is sufficient to obtain a period of cosmological inflation that
the {\it slow-roll conditions} for $\varphi$ are satisfied. Recall that
the equation of motion for a homogeneous scalar field in a cosmological
space-time is (as follows from (\ref{scalarlag})) is
\begin{equation} \label{scalareom}
{\ddot \varphi} + 3 H {\dot \varphi} \, = \, - V^{\prime}(\varphi) \, ,
\end{equation}
where a prime indicates the derivative with respect to $\varphi$. In order
that the scalar field roll slowly, it is necessary that
\begin{equation} \label{roll}
{\ddot \varphi} \, \ll \, 3 H {\dot \varphi}
\end{equation}
such that the first term in the scalar field equation of motion 
(\ref{scalareom}) is negligible. In this case, the condition (\ref{tempcond})
becomes
\begin{equation} \label{SRcond1}
({{V^{\prime}} \over V})^2 \, \ll \, 48 \pi G \,
\end{equation}
and (\ref{roll}) becomes
\begin{equation} \label{SRcond2}
{{V^{\prime \prime}} \over V} \, \ll \, 24 \pi G \, .
\end{equation}

In the initial model of inflation using scalar fields (``old inflation'' 
\cite{Guth}), it was assumed that $\varphi$ was initially in a false
vacuum with large potential energy. Hence, the conditions for inflation
are trivially satisfied. To end inflation, a quantum tunneling event from
the false vacuum to the true vacuum \cite{Coleman} was invoked (see e.g.
\cite{RMP} for a pedagogical review). This model, however, has a graceful 
exit problem since the tunneling leads to an initially microscopical bubble
of the true vacuum which cannot grow to encompass our presently observed 
universe - the flatness problem of SBB cosmology in a new form. Hence,
attention shifted to models in which the scalar field $\varphi$ is slowly
rolling during inflation.

There are many models of scalar field-driven inflation. Many of them can be
divided into three groups \cite{Kolb}: small-field inflation, large-field 
inflation and hybrid inflation. {\it Small-field inflationary models} are based
on ideas from spontaneous symmetry breaking in particle physics. We take the
scalar field to have a potential of the form
\begin{equation} \label{Mexican}
V(\varphi) \, = \, {1 \over 4} \lambda (\varphi^2 - \sigma^2)^2 \, ,
\end{equation}
where $\sigma$ can be interpreted as a symmetry breaking scale, and
$\lambda$ is a dimensionless coupling constant. The hope of initial
small-field models (``new inflation'' \cite{new}) was that the scalar
field would begin rolling close to its symmetric point $\varphi = 0$,
where thermal equilibrium initial conditions would localize it in the
early universe. At sufficiently high temperatures, $\varphi = 0$ 
is a stable ground state of the one-loop finite temperature effective 
potential $V_T(\varphi)$ (see e.g. \cite{RMP} for a review). 
Once the temperature drops to a value 
smaller than the critical temperature $T_c$, $\varphi = 0$ turns into 
an unstable local maximum of $V_T(\varphi)$, and $\varphi$ is free to 
roll towards a ground state of the zero temperature potential (\ref{Mexican}). 
The direction of the initial rolling is triggered by quantum fluctuations. 
The reader can
easily check that for the potential (\ref{Mexican}) the slow-roll conditions
cannot be satisfied if $\sigma \ll m_{pl}$, where $m_{pl}$ is the
Planck mass which is related to $G$. 
If the potential is modified to a Coleman-Weinberg \cite{CW} form 
\begin{equation} \label{CWpot}
V(\varphi) \, = \, {{\lambda} \over 4} \varphi^4
\bigl[ {\rm ln}  {{|\varphi|} \over {\sigma}} - {1 \over 4} \bigr]
+ {1 \over {16}} \lambda \sigma^4
%+ {{\lambda^2} \over {44 \pi^2}} \varphi^4 
%\bigl[ {\rm ln} ({{\varphi^2} \over {M^2}}) - {{25} \over 6} \bigr]
\end{equation}
(where $\sigma$ denotes the value of the minimum of the potential) 
then the slow-roll conditions can be satisfied. However, this
corresponds to a severe fine-tuning of the shape of the potential.
A further problem for most small-field models of inflation (see 
e.g. \cite{Goldwirth} for a review) is that the slow-roll trajectory
is not an attractor in phase space. In order to end up close to the
slow-roll trajectory, the initial field velocity must be constrained
to be very small. This {\it initial condition problem} of small-field
models of inflation effects a number of recently proposed brane inflation
scenarios, see e.g. \cite{GhazalScott} for a discussion. 

There is another reason for abandoning small-field inflation models: in
order to obtain a sufficiently small amplitude of density fluctuations,
the interaction coefficients of $\varphi$ must be very small (this
problem is discussed in detail at the beginning of Section 4). In
particular, this makes it inconsistent to assume that $\varphi$ started
out in thermal equilibrium. In the absence of thermal equilibrium, the
phase space of initial conditions is much larger for large values of
$\varphi$. 

This brings us to the discussion of large-field inflation
models, initially proposed in \cite{chaotic} under the name
``chaotic inflation''. The simplest example is
provided by a massive scalar field with potential
\begin{equation}
V(\varphi) \, = \, {1 \over 2} m^2 \varphi^2 \, ,
\end{equation}
where $m$ is the mass. It is assumed that the scalar field rolls towards
the origin from large values of $|\varphi|$. It is a simple exercise 
for the reader to verify that the slow-roll conditions (\ref{SRcond1}) and
(\ref{SRcond2}) are satisfied provided
\begin{equation}
|\varphi| \, > \, {1 \over {\sqrt{12 \pi}}} m_{pl} \, .
\end{equation}
Values of $|\varphi|$ comparable or greater than $m_{pl}$ are also
required in other realizations of large-field inflation. Hence, one
may worry whether such a toy model can consistently be embedded in
a realistic particle physics model, e.g. supergravity. In many such
models $V(\varphi)$ receives supergravity-induced correction terms
which destroy the flatness of the potential for $|\varphi| > m_{pl}$.
However, as discussed e.g. in \cite{Linderev}, if the flatness
of the potential is protected by some symmetry, then it can survive
inclusion of the correction terms. As will be discussed later, a value
of $m \sim 10^{13}$GeV is required in order to obtain the observed
amplitude of density fluctuations. Hence, the configuration space of
field values with $|\varphi| > m_{pl}$ but $V(\varphi) < m_{pl}^4$ is
huge. It can also be verified that the slow-roll trajectory is a local
attractor in field initial condition space \cite{Kung}, even including
metric fluctuations at the perturbative level \cite{Feldman}.

With two scalar fields it is possible to construct a class of models
which combine some of the nice features of large-field inflation
(large phase space of initial conditions yielding inflation) and of
small-field inflation (inflation taking place at sub-Planckian field
values). These are models of hybrid inflation \cite{hybrid}. To give a prototypical
example, consider two scalar fields $\varphi$ and $\chi$ with a potential
\begin{equation} \label{hybridpot}
V(\varphi, \chi) \, = \, {1 \over 4} \lambda_{\chi} (\chi^2 - \sigma^2)^2
+ {1 \over 2} m^2 \varphi^2 - {1 \over 2} g^2 \varphi^2 \chi^2 \, .
\end{equation}
In the absence of thermal equilibrium, it is natural to assume that 
$|\varphi|$ begins at large values, values for which the effective mass
of $\chi$ is positive and hence $\chi$ begins at $\chi = 0$. The parameters
in the potential (\ref{hybridpot}) are now chosen such that $\varphi$
is slowly rolling for values of $|\varphi|$ somewhat smaller than $m_{pl}$,
but that the potential energy for these field values is dominated by
the first term on the right-hand side of (\ref{hybridpot}). The reader
can easily verify that for this model it is no longer required to have
values of $|\varphi|$ greater than $m_{pl}$ in order to obtain slow-rolling
\footnote{Note that the slow-roll conditions (\ref{SRcond1}) and (\ref{SRcond2})
were derived assuming that $H$ is given by the contribution of
$\varphi$ to $V$ which is not the case here.}
The field $\varphi$ is slowly rolling whereas the potential energy is
determined by the contribution from $\chi$. Once $|\varphi|$ drops to the
value
\begin{equation}
|\varphi_c| \, = \, {{{\sqrt{\lambda_{\chi}}}} \over g} \sigma 
\end{equation}
the configuration $\chi = 0$ becomes unstable and decays to its ground
state $|\chi| = \sigma$, yielding a graceful exit from inflation.
Since in this example the ground state of $\chi$ is not unique, there
is the possibility of the formation of topological defects at the end
of inflation (see \cite{ShellVil,HK,RHBrev0} for reviews of topological
defects in cosmology, and the lectures by Polchinski
\cite{PolCar} for a discussion of how this
scenario arises in brane inflation models).

After the slow-roll conditions break down, the period of inflation ends,
and the inflaton begins to oscillate around its ground state.
Due to couplings of $\varphi$ to other matter fields, the energy of the
universe, which at the end of the period of inflation is stored completely
in $\varphi$, gets transferred to the matter fields of the particle
physics Standard Model. Initially, the energy transfer was described
perturbatively \cite{DolLin,AFW}. Later, it was realized \cite{TB,KLS,STB,KLS2}
that through a parametric resonance instability, particles are very
rapidly produced, leading to a fast energy transfer (``preheating'').
The quanta later thermalize, and thereafter the universe evolves as described
by SBB cosmology.

\section{Theory of Cosmological Perturbations: A Short Review}

After this review of inflationary cosmology (see e.g. \cite{Lythbook} for
a more complete recent review), we turn to the discussion of the main
success of inflationary cosmology, namely the fact that it provides
a causal mechanism for generating small amplitude inhomogeneities. The reader is
referred to \cite{MFB} for a comprehensive analysis of this theory of
cosmological perturbations, and \cite{RHBrev2} for a pedagogical overview,
from which most of this section is drawn.

First, we describe the Newtonian theory of cosmological perturbations,
mainly to develop intuition for the main physical effects. The range
of validity of the Newtonian analysis is restricted to sub-Hubble
scales at late times. In the second subsection, we then summarize
the general relativistic theory of fluctuations. Linear fluctuations
can be trivially quantized. The quantization of cosmological perturbations
is presented in Subsection 3.

\subsection{Newtonian Theory of Cosmological Perturbations}
\label{rhbsec:1}

The growth of density fluctuations is a consequence of the
purely attractive nature of the gravitational force. Given 
an density excess $\delta \rho$ localized about some point ${\bf x}$ in space.
This inhomogeneity produces a force which attracts
the surrounding matter towards ${\bf x}$. The magnitude of this
force is proportional to $\delta \rho$. Hence, in a non-expanding
background, by Newton's second law
\begin{equation} \label{rhbeq1}
\ddot{\delta \rho} \, \sim \, G \delta \rho \, ,
\end{equation}
where $G$ is Newton's gravitational constant. Thus, an
exponential growth of the fluctuations is induced. 

If, as required by consistency in General Relativity, we
consider density fluctuations in an expanding background,
then the expansion of space leads to a friction term in (\ref{rhbeq1}).
Hence, instead of an exponential instability to the development of
fluctuations, the growth of fluctuations
will be as a power of time. The main goal of the theory
of cosmological perturbations is to determine how the power-law
instability depends on the background cosmology
and on the length scale of the fluctuations.

\subsection{Perturbations about Minkowski Space-Time}

To develop some physical intuition, we first consider the
evolution of hydrodynamical matter fluctuations in a fixed
non-expanding background.  

In this context, matter is described
by a perfect fluid, and gravity by the Newtonian gravitational
potential $\varphi$. The fluid variables are the energy density
$\rho$, the pressure $p$, the fluid velocity ${\bf v}$, and
the entropy density $S$. The basic hydrodynamical equations
are
\begin{eqnarray} \label{rhbeq2}
\dot{\rho} + \nabla_p \cdot (\rho {\bf v}) & = & 0 \nonumber \\
\dot{{\bf v}} + ({\bf v} \cdot \nabla_p) {\bf v} + {1 \over {\rho}} \nabla_p p
+ \nabla_p \varphi & = & 0 \nonumber \\
\nabla_p^2 \varphi & = & 4 \pi G \rho \\
\dot{S} + ({\bf v} \cdot \nabla_p) S & = & 0 \nonumber \\
p & = & p(\rho, S) \, , \nonumber
\end{eqnarray}
where the subscript $p$ indicates that physical as opposed to
comoving coordinates are used.
The first equation is the continuity equation, the second is the Euler
(force) equation, the third is the Poisson equation of Newtonian gravity,
the fourth expresses entropy conservation, and the last describes
the equation of state of matter. The derivative with respect to time
is denoted by an over-dot.

The background is given by the background energy density $\rho_o$, 
the background
pressure $p_0$, vanishing velocity, constant gravitational potential
$\varphi_0$ and constant entropy density $S_0$. Note that the
background Poisson equation is not satisfied.

The equations for cosmological perturbations are obtained by perturbing
the fluid variables about the background,
\begin{eqnarray} \label{rhbeq3}
\rho & = & \rho_0 + \delta \rho \nonumber \\
{\bf v} & = & \delta {\bf v} \nonumber \\
p & = & p_0 + \delta p \\
\varphi & = & \varphi_0 + \delta \varphi \nonumber \\
S & = & S_0 + \delta S \, , \nonumber
\end{eqnarray}
where the fluctuating fields $\delta \rho, \delta {\bf v}, \delta p,
\delta \varphi$ and $\delta S$ are functions of space and time, by
inserting these expressions into the basic hydrodynamical equations
(\ref{rhbeq2}), and by linearizing. After combining the resulting first
order equations, we get the following
differential equations for the energy density fluctuation $\delta \rho$
and the entropy perturbation $\delta S$
\begin{eqnarray} \label{rhbeq4}
\ddot{\delta \rho} - c_s^2 \nabla_p^2 \delta \rho - 4 \pi G \rho_0 \delta \rho
& = & \sigma \nabla_p^2 \delta S \\
\dot \delta S \ & = & 0 \, , \nonumber
\end{eqnarray}
where the variables $c_s^2$ and $\sigma$ describe the equation of state
\begin{equation} \label{pressurepert}
\delta p \, = \, c_s^2 \delta \rho + \sigma \delta S
\end{equation}
with
\begin{equation}
c_s^2 \, = \, \bigl({{\delta p} \over {\delta \rho}}\bigr)_{|_{S}}
\end{equation}
denoting the square of the speed of sound.

Since the equations are linear, we can work in Fourier space. Each
Fourier component $\delta \rho_k(t)$ of the fluctuation field 
$\delta \rho({\bf x}, t)$ 
\begin{equation}
\delta \rho ({\bf x}, t) \, = \, 
\int e^{i {\bf k} \cdot {\bf x}} \delta \rho_k(t)
\end{equation}
evolves independently.

The fluctuations can be classified as follows: If 
$\delta S$ vanishes, we have {\bf adiabatic} fluctuations. If
the $\delta S$ is non-vanishing but 
$\dot{\delta \rho} = 0$, we speak of an {\bf entropy} fluctuation.

The first conclusions we can draw from the basic perturbation
equations (\ref{rhbeq4}) are that \\
1) entropy fluctuations do not grow, \\
2) adiabatic fluctuations are time-dependent, and \\
3) entropy fluctuations seed an adiabatic mode.

Taking a closer look at the equation of motion for
$\delta \rho$, we see that the third term on the left hand side
represents the force due to gravity, a purely attractive force 
yielding an instability of flat space-time to the development of
density fluctuations (as discussed earlier, see (\ref{rhbeq1})).
The second term on the left hand side of (\ref{rhbeq4}) represents
a force due to the fluid pressure which tends to set up pressure waves.
In the absence of entropy fluctuations, the evolution of $\delta \rho$
is governed by the combined action of both pressure and gravitational
forces.

Restricting our attention to adiabatic fluctuations, we see from
(\ref{rhbeq4}) that there is a critical wavelength, the Jeans length,
whose wavenumber $k_J$ is (in physical coordinates) given by
\begin{equation} \label{Jeans}
k_J \, = \, \bigl({{4 \pi G \rho_0} \over {c_s^2}}\bigr)^{1/2} \, .
\end{equation}
Fluctuations with wavelength longer than the Jeans length ($k \ll k_J$)
grow exponentially
\begin{equation} \label{expgrowth}
\delta \rho_k(t) \, \sim \, e^{\omega_k t} \,\, {\rm with} \,\,
\omega_k \sim 4 (\pi G \rho_0)^{1/2}
\end{equation}
whereas short wavelength modes ($k \gg k_J$) oscillate with 
frequency $\omega_k \sim c_s k$. Note that the value of the
Jeans length depends on the equation of state of the background.
For a background dominated by relativistic radiation, the Jeans
length is large (of the order of the Hubble radius $H^{-1}(t)$), 
whereas for pressure-less matter it goes to zero.

Next, we study Newtonian
cosmological fluctuations about an expanding background. In this
case, the background equations are consistent (the non-vanishing
average energy density leads to cosmological expansion). However,
we are neglecting general relativistic effects (the 
fluctuations of the metric) which
dominante on length scales larger than the Hubble
radius $H^{-1}(t)$. 

The background cosmological model is given by the energy density
$\rho_0(t)$, the pressure $p_0(t)$, and the recessional velocity
${\bf v}_0 = H(t) {\bf x}_p$ where ${\bf x}_p$ is the physical 
coordinate vector. The space- and time-dependent
fluctuating fields are defined in analogy to the previous section:
\begin{eqnarray} \label{fluctansatz2}
\rho(t, {\bf x}) & = & \rho_0(t) \bigl(1 + \delta_{\epsilon}(t, {\bf x}) 
\bigr)\nonumber \\
{\bf v}(t, {\bf x}) & = & {\bf v}_0(t, {\bf x}) + \delta {\bf v}(t, {\bf x}) 
\\
p(t, {\bf x}) & = & p_0(t) + \delta p(t, {\bf x}) \, , \nonumber 
\end{eqnarray}
where $\delta_{\epsilon}$ is the fractional energy density perturbation
(we are interested in the fractional rather than in the absolute energy
density fluctuation!), and the pressure perturbation $\delta p$ is
defined as in (\ref{pressurepert}). In addition, there is the
possibility of a non-vanishing entropy perturbation defined as in
(\ref{rhbeq3}).

We now insert this ansatz into the basic hydrodynamical equations 
(\ref{rhbeq2}), linearize in the perturbation variables, and combine
the first order differential equations 
for $\delta_{\epsilon}$ and $\delta p$ into a single second order
differential equation for $\delta_{\epsilon}$. The result simplifies
if we work in comoving coordinates $\bf x$. After some algebra, we obtain 
the following equation
which describes the time evolution of density fluctuations:
\begin{equation} \label{Newtoneq}
\ddot{\delta_{\epsilon}} + 2 H \dot{\delta_{\epsilon}} 
- {{c_s^2} \over {a^2}} \nabla^2 \delta_{\epsilon} 
- 4 \pi G \rho_0 \delta_{\epsilon} \, 
= \, {{\sigma} \over {\rho_0 a^2}} \delta S \, ,
\end{equation}
where $\nabla$ is the partial derivative vector with respect to
comoving coordinates.
In addition, we have the equation of entropy conservation
\begin{equation}
\dot{\delta S} \, = \, 0 \, .
\end{equation}

Comparing with the equations (\ref{rhbeq4}) obtained in the absence of
an expanding background, we see that the only difference is the presence
of a Hubble damping term in the equation for $\delta_{\epsilon}$. This
term will moderate the exponential instability of the background to
long wavelength density fluctuations. In addition, it will lead to a
damping of the oscillating solutions on short wavelengths. More specifically,
for physical wavenumbers $k_p \ll k_J$ (where $k_J$ is again given by
(\ref{Jeans})), and in a matter-dominated background cosmology, the
general solution of (\ref{Newtoneq}) in the absence of any entropy
fluctuations is given by
\begin{equation} \label{Newtonsol}
\delta_k(t) \, = \, c_1 t^{2/3} + c_2 t^{-1} \, ,
\end{equation}
where $c_1$ and $c_2$ are two constants determined by the initial
conditions, and we have dropped the subscript $\epsilon$ in expressions
involving $\delta_{\epsilon}$. 
There are two fundamental solutions, the first a
growing mode with $\delta_k(t) \sim a(t)$, the second a decaying
mode with $\delta_k(t) \sim t^{-1}$.
On short wavelength, one obtains damped oscillatory motion:
\begin{equation} \label{Newtonsolosc}
\delta_k(t) \, \sim \, a^{-1/2}(t) exp \bigl( \pm i c_s k \int dt' a^{-1}(t')
\bigr) \, .
\end{equation}

Before going on to the relativitic theory of cosmological perturbations,
we will pause to introduce terminology used in cosmology to describe the
fluctuations.

Let us consider perturbations on a fixed
comoving length scale given by a comoving wavenumber $k$. 
The corresponding physical length increases
as $a(t)$. This is to be compared to the Hubble radius $H^{-1}(t)$
which scales as $t$ provided $a(t)$ grows as a power of $t$. In
the late time Universe, $a(t) \sim t^{1/2}$ in the radiation-dominated
phase (i.e. for $t < t_{eq}$), and $a(t) \sim t^{2/3}$ in the 
matter-dominated period ($t_{eq} < t < t_0$). 
Thus, at sufficiently early times, all comoving scales
had a physical length larger than the Hubble radius. If we consider 
large cosmological scales (e.g. those corresponding to the observed
CMB anisotropies or to galaxy clusters), the time $t_H(k)$ of 
``Hubble radius crossing'' (when the physical length was equal to the
Hubble radius) was in fact later than $t_{eq}$. The time of Hubble 
radius crossing plays an
important role in the evolution of cosmological perturbations.

Cosmological fluctuations can be described either in position space
or in momentum space. In position space, we compute the root mean
square mass fluctuation $\delta M / M(k, t)$ in a sphere of radius
$l = 2 \pi / k$ at time $t$. A scale-invariant spectrum of fluctuations is
defined by the relation
\begin{equation} \label{scaleinv}
{{\delta M} \over M}(k, t_H(k)) \, = \, {\rm const.} \, .
\end{equation}
Such a spectrum was first suggested by Harrison \cite{Harrison}
and Zeldovich \cite{Zeldovich} as a reasonable choice for the
spectrum of cosmological fluctuations. The ``spectral
index'' $n$ of cosmological fluctuations is defiined by the relation
\begin{equation} \label{specindex}
\bigl({{\delta M} \over M}\bigr)^2(k, t_H(k)) \, \sim \, k^{n - 1} \, .
\end{equation}
Thus, a scale-invariant spectrum corresponds to $n = 1$.

To go to momentum space
representation, the fractional spatial density contrast
is expanded in a Fourier series:
\begin{equation} \label{Fourier}
\delta_{\epsilon}({\bf x}, t) \, = \, 
\int d^3k {\tilde{\delta_{\epsilon}}}({\bf k}, t) e^{i {\bf k} \cdot {\bf x}}
\, .
\end{equation}
The {\bf power spectrum} $P_{\delta}$ of density fluctuations is defined by
\begin{equation} \label{densspec}
P_{\delta}(k) \, = \, {1 \over {2 \pi^2}}
k^3 |{\tilde{\delta_{\epsilon}}}(k)|^2 \, ,
\end{equation}
where $k$ is the magnitude of ${\bf k}$. For simplicity, the
distribution of fluctuations is taken to be Gaussian so that the  
fluctuation amplitude only depends on $k$.

The power spectrum of the gravitational potential $\varphi$
is defined by
\begin{equation} \label{gravspec}
P_{\varphi}(k) \, = \, {1 \over {2 \pi^2}}
k^3 |{\tilde{\delta \varphi}}(k)|^2 \, .
\end{equation}
The two power spectra (\ref{densspec}) and (\ref{gravspec})
are related by the Poisson equation (\ref{rhbeq2})
\begin{equation} \label{relspec}
P_{\varphi}(k) \, \sim \, k^{-4} P_{\delta}(k) \, .
\end{equation}

The condition of scale-invariance can be expressed in terms of the power spectrum evaluated at a fixed time. To obtain
this condition, we first use the time dependence of the 
fractional density fluctuation from (\ref{Newtonsol}) to determine
the mass fluctuations at a fixed time. We need to use
the fact that the time of Hubble radius crossing is given by
\begin{equation} \label{Hubble}
a(t_H(k)) k^{-1} \, = \, \beta t_H(k) \, ,
\end{equation} 
where $\beta = 2$ or $\beta = 3/2$ in the radiation and matter
dominated phases, respectively. Making use of (\ref{specindex})
we find 
\begin{equation} \label{spec2}
\bigl({{\delta M} \over M}\bigr)^2(k, t) \, \sim \, k^{n + 3} \, .
\end{equation}
Since, for reasonable values of the index of the power spectrum, 
$\delta M / M (k, t)$ is dominated by the Fourier modes with
wavenumber $k$, we find that (\ref{spec2}) implies
\begin{equation} \label{spec3}
P_{\delta}(k) \, \sim \, k^{n + 3} \, ,
\end{equation}
or, equivalently,
\begin{equation} \label{spec4}
P_\varphi(k) \, \sim \, k^{n - 1} \, .
\end{equation}

\subsection{Relativistic Theory of Cosmological Fluctuations}

%\subsection{Introduction}

The Newtonian theory of cosmological fluctuations discussed in the
previous section breaks down on scales larger than the Hubble radius
because it neglects perturbations of the metric, and because on large
scales the metric fluctuations dominate the dynamics.

To show why metric fluctuations are important on scales larger than the 
Hubble radius, we can use a ``separate universe" argument.  On such
scales, one should be able to approximately describe the
evolution of the space-time by applying the first 
FRW equation (\ref{FRW1}) of homogeneous
and isotropic cosmology to the local Universe (this approximation is
made more rigorous in \cite{Afshordi}).
Based on this equation, a large-scale fluctuation of the
energy density will lead to a fluctuation (``$\delta a$'') of
the scale factor $a$ which grows in time, a manifestation of the
self gravitational amplification of fluctuations on length scales $\lambda$
greater than the Hubble radius.

Let us now turn to the rigorous analysis of cosmological
fluctuations in the context of general relativity, where both metric
and matter inhomogeneities are taken into account. We will consider
fluctuations about a homogeneous and isotropic background cosmology,
given by the metric (\ref{background}), which can be written in
conformal time $\eta$ (defined by $dt = a(t) d\eta$) as
\begin{equation} \label{background2}
ds^2 \, = \, a(\eta)^2 \bigl( d\eta^2 - d{\bf x}^2 \bigr) \, .
\end{equation}

The theory of cosmological perturbations is based on expanding the Einstein
equations to linear order about the background metric. The theory was
initially developed in pioneering works by Lifshitz \cite{Lifshitz}. 
Significant progress in the understanding of the physics of cosmological
fluctuations was achieved by Bardeen \cite{Bardeen} who realized the
importance of subtracting gauge artifacts (see below) from the
analysis (see also \cite{PV}). The following discussion is based on
Part I of the comprehensive review article \cite{MFB}. Other reviews -
emphasizing different aspects or approaches - are 
\cite{Kodama,Ellis,Hwang,Durrer}.

%\subsection{Classifying Fluctuations}

The first step in the analysis of metric fluctuations is to
classify them according to their transformation properties
under spatial rotations. There are scalar, vector and second rank
tensor fluctuations. In linear theory, there is no coupling
between the different fluctuation modes, and hence they evolve
independently (for some subtleties in this classification, see
\cite{Stewart}). 

We begin by expanding the metric about the FRW background metric
$g_{\mu \nu}^{(0)}$ given by (\ref{background2}):
\begin{equation} \label{pertansatz}
g_{\mu \nu} \, = \, g_{\mu \nu}^{(0)} + \delta g_{\mu \nu} \, .
\end{equation}
The background metric depends only on time, whereas the metric
fluctuations $\delta g_{\mu \nu}$ depend on both space and time.
Since the metric is a symmetric tensor, there are at first sight
10 fluctuating degrees of freedom in $\delta g_{\mu \nu}$.

There are four degrees of freedom which correspond to scalar metric
fluctuations (the only four ways of constructing a metric from
scalar functions):
\begin{equation} \label{scalarfl}
\delta g_{\mu \nu} \, = \, a^2 \left(
\begin{array} {cc}
2 \phi & -B_{,i} \\
-B_{,i} & 2\bigl(\psi \delta_{ij} - E_{,ij} \bigr) 
\end{array}
\right) \, ,
\end{equation}
where the four fluctuating degrees of freedom are denoted (following
the notation of \cite{MFB}) $\phi, B, E$, and $\psi$, a comma denotes
the ordinary partial derivative (if we had included spatial curvature
of the background metric, it would have been the covariant derivative
with respect to the background spatial metric), and 
$\delta_{ij}$ is the Kronecker symbol.

There are four vector degrees of freedom of metric fluctuations,
consisting of the four ways of constructing metric fluctuations from
three vectors:
\begin{equation} \label{vectorfl}
\delta g_{\mu \nu} \, = \, a^2 \left(
\begin{array} {cc}
0 & -S_i \\
-S_i & F_{i,j} + F_{j,i} 
\end{array}
\right) \, ,
\end{equation}
where $S_i$ and $F_i$ are two divergence-less vectors (for a vector
with non-vanishing divergence, the divergence contributes to the
scalar gravitational fluctuation modes).

Finally, there are two tensor modes which correspond to the two
polarization states of gravitational waves:
\begin{equation} \label{tensorfl}
\delta g_{\mu \nu} \, = \, -a^2 \left(
\begin{array} {cc}
0 & 0 \\
0 & h_{ij} 
\end{array}
\right) \, ,
\end{equation}
where $h_{ij}$ is trace-free and divergence-less
\begin{equation}
h_i^i \, = \, h_{ij}^j \, = \, 0 \, .
\end{equation}

Gravitational waves do not couple at linear order to the matter
fluctuations. Vector fluctuations decay in an expanding background
cosmology and hence are not usually cosmologically important.
Thus, the most important fluctuations, at least in inflationary cosmology,
are the scalar metric fluctuations, the fluctuations which couple
to matter inhomogeneities and which are the relativistic generalization
of the Newtonian perturbations considered in the previous section.

%\subsection{Gauge Transformation}

The theory of cosmological perturbations is at first sight complicated
by the issue of gauge invariance. The coordinates $t, {\bf x}$ of space-time 
carry no
independent physical meaning. By performing a small-amplitude 
transformation of the space-time coordinates
(called ``gauge transformation'' in the following), we can 
create ``fictitious'' fluctuations in a homogeneous and isotropic
Universe. These modes are ``gauge artifacts''.

In the following we take an ``active'' view of gauge transformation.
Consider two space-time manifolds, one of them a homogeneous
and isotropic Universe ${\cal M}_0$, the other a physical Universe 
${\cal M}$ with inhomogeneities. A choice of coordinates can be considered
as a mapping ${\cal D}$ between the manifolds ${\cal M}_0$ and ${\cal M}$.
A second mapping ${\tilde{\cal D}}$ will map the
same point in ${\cal M}_0$
into a different point in ${\cal M}$. Using the inverse of these maps
${\cal D}$ and ${\tilde{\cal D}}$, we can assign two different sets of
coordinates to points in ${\cal M}$. 

Consider now a physical quantity $Q$ (e.g. the Ricci scalar)
on ${\cal M}$, and the corresponding
physical quantity $Q^{(0)}$ on ${\cal M}_0$ Then, in the first coordinate
system given by the mapping ${\cal D}$, the perturbation $\delta Q$ of
$Q$ at the point $p \in {\cal M}$ is defined by
\begin{equation}
\delta Q(p) \, = \, Q(p) - Q^{(0)}\bigl({\cal D}^{-1}(p) \bigr) \, .
\end{equation}
In the second coordinate system given by ${\tilde{\cal D}}$,
the perturbation is defined by
\begin{equation}
{\tilde{\delta Q}}(p) \, = \, Q(p) - 
Q^{(0)}\bigl({\tilde{{\cal D}}}^{-1}(p) \bigr) \, .
\end{equation}
The difference
\begin{equation}
\Delta Q(p) \, = \, {\tilde{\delta Q}}(p) - \delta Q(p)
\end{equation}
is a gauge artifact and carries no physical significance.

Some of the metric perturbation degrees of freedom introduced in
the first subsection are thus gauge artifacts. To isolate these, 
we must study how coordinate transformations act on the metric.
There are four independent gauge degrees of freedom corresponding
to the coordinate transformations
\begin{equation}
x^{\mu} \, \rightarrow \, {\tilde x}^{\mu} = x^{\mu} + \xi^{\mu} \, .
\end{equation}
The time component $\xi^0$ of $\xi^{\mu}$ leads to a scalar metric
fluctuation. The spatial three vector $\xi^i$ can be decomposed as
\begin{equation}
\xi^i \, = \, \xi^i_{tr} + \gamma^{ij} \xi_{,j}
\end{equation}
(where $\gamma^{ij}$ is the spatial background metric)
into a transverse piece $\xi^i_{tr}$ which has two degrees of freedom
which yield vector perturbations, and the second term (given by
the gradient of a scalar $\xi$) which gives
a scalar fluctuation. Thus, there are
two scalar gauge modes given by $\xi^0$ and $\xi$, and two vector
modes given by the transverse three vector $\xi^i_{tr}$. Thus,
there remain two physical scalar and two vector
fluctuation modes. The gravitational waves are gauge-invariant. 

Let us now focus on how the scalar gauge transformations (i.e. the
transformations given by $\xi^0$ and $\xi$) act on the scalar
metric fluctuation variables $\phi, B, E$, and $\psi$. An immediate
calculation yields:
\begin{eqnarray}
{\tilde \phi} \, &=& \, \phi - {{a'} \over a} \xi^0 - (\xi^0)^{'} \nonumber \\
{\tilde B} \, &=& \, B + \xi^0 - \xi^{'} \\
{\tilde E} \, &=& \, E - \xi \nonumber \\
{\tilde \psi} \, &=& \, \psi + {{a'} \over a} \xi^0 \, , \nonumber
\end{eqnarray}
where a prime indicates the derivative with respect to conformal time $\eta$.

There are two approaches to deal with the gauge ambiguities. The first is
to fix a gauge, i.e. to pick conditions on the coordinates which
completely eliminate the gauge freedom, the second is to work with a
basis of gauge-invariant variables.

If one wants to adopt the gauge-fixed approach, there are many
different gauge choices. Note that the often used synchronous gauge
determined by $\delta g^{0 \mu} = 0$ does not totally fix the
gauge. A convenient system which completely fixes the coordinates
is the so-called {\bf longitudinal} or {\bf conformal Newtonian gauge}
defined by $B = E = 0$.

If one prefers a gauge-invariant approach, there are many choices
of gauge-invariant variables. A convenient basis first introduced
by \cite{Bardeen} is the basis $\Phi, \Psi$ given by 
\begin{eqnarray} \label{givar}
\Phi \, &=& \, \phi + {1 \over a} \bigl[ (B - E')a \bigr]^{'} \\
\Psi \, &=& \, \psi - {{a'} \over a} (B - E') \, .
\end{eqnarray}
The gauge-invariant
variables $\Phi$ and $\Psi$ coincide with the corresponding
diagonal metric perturbations $\phi$ and $\psi$ in longitudinal
gauge. 

Note that the variables defined in (\ref{givar}) are gauge-invariant only
under linear space-time coordinate transformations. Beyond
linear order, the structure of perturbation theory becomes much
more involved. In fact, one can show \cite{SteWa} that the only
fluctuation variables which are invariant under all coordinate
transformations are perturbations of variables which are constant
in the background space-time. Beyond linear order there is also
mixing of scalar, vector and tensor modes.

%\subsection{Equations of Motion}

To derive the equations of motion for the fluctuations, the
starting point is the set of Einstein equations
\begin{equation} \label{Einstein}
G_{\mu\nu} \, = \, 8 \pi G T_{\mu\nu} \, , 
\end{equation}
where $G_{\mu\nu}$ is the Einstein tensor associated with the space-time
metric $g_{\mu\nu}$, and $T_{\mu\nu}$ is the energy-momentum tensor of matter.
We insert the ansatz for metric and matter perturbed about a FRW 
background $\bigl(g^{(0)}_{\mu\nu}(\eta) ,\, \varphi_0(\eta)\bigr)$:
\begin{eqnarray} \label{pertansatz2}
g_{\mu\nu} ({\bf x}, \eta) & = & g^{(0)}_{\mu\nu} (\eta) + \delta g_{\mu\nu}
({\bf x}, \eta) \\
\varphi ({\bf x}, \eta) & = & \varphi_0 (\eta) + \delta \varphi
({\bf x}, \eta) \, , 
\end{eqnarray}
(where we have for simplicity replaced general matter by a scalar
matter field $\varphi$)
and expand to linear order in the fluctuating fields, obtaining the
following equations:
\begin{equation} \label{linein}
\delta G_{\mu\nu} \> = \> 8 \pi G \delta T_{\mu\nu} \, .
\end{equation}
In the above, $\delta g_{\mu\nu}$ is the perturbation in the metric and $\delta
\varphi$ is the fluctuation of the matter field $\varphi$.

In a gauge-fixed approach, one can start with the metric in
longitudinal gauge
\begin{equation} \label{longit}
ds^2 \, = \, a^2 \bigl[(1 + 2 \phi) d\eta^2
- (1 - 2 \psi)\gamma_{ij} dx^i dx^j \bigr] \,
\end{equation}
and insert this ansatz into the general perturbation equations
(\ref{linein}). This yields the following set of equations of motion:
\begin{eqnarray} \label{perteom1}
- 3 {\cal H} \bigl( {\cal H} \phi + \psi^{'} \bigr) + \nabla^2 \psi \,
&=& \, 4 \pi G a^2 \delta T^{0}_0 \nonumber \\
\bigl( {\cal H} \phi + \psi^{'} \bigr)_{, i} \,
&=& 4 \pi G a^2 \delta T^{0}_i \\
\bigl[ \bigl( 2 {\cal H}^{'} + {\cal H}^2 \bigr) \phi + {\cal H} \phi^{'}
+ \psi^{''} + 2 {\cal H} \psi^{'} \bigr] \delta^i_j && \nonumber \\
+ {1 \over 2} \nabla^2 D \delta^i_j - {1 \over 2} \gamma^{ik} D_{, kj} \,
&=& - 4 \pi G a^2 \delta T^{i}_j \, , \nonumber
\end{eqnarray}
where $D \equiv \phi - \psi$ and ${\cal H} = a'/a$. 

If no anisotropic stress
is present in the matter at linear order in fluctuating fields, i.e.
if $\delta T^i_j = 0$ for $i \neq j$, then the two metric fluctuation
variables coincide, i.e. $\phi \, = \, \psi$ .
This will be the case in most simple cosmological models, e.g. in
theories with matter described by a set of scalar fields with
canonical form of the action, and in the case of a perfect fluid
with no anisotropic stress.

In the simple case in which matter described
in terms of a single scalar field $\varphi$, 
then in longitudinal gauge
(\ref{perteom1}) reduce to the following
set of equations of motion 
\begin{eqnarray} \label{perteom2}
\nabla^2 \phi - 3 {\cal H} \phi^{'} - 
\bigl( {\cal H}^{'} + 2 {\cal H}^2 \bigr) \phi \, &=& \, 
4 \pi G \bigl( \varphi^{'}_0 \delta \varphi^{'} + 
V^{'} a^2 \delta \varphi \bigr) \nonumber \\
\phi^{'} + {\cal H} \phi \, &=& \, 4 \pi G \varphi^{'}_0 \delta \varphi \\
\phi^{''} + 3 {\cal H} \phi^{'} + 
\bigl( {\cal H}^{'} + 2 {\cal H}^2 \bigr) \phi \, &=& \, 
4 \pi G \bigl( \varphi^{'}_0 \delta \varphi^{'} - 
V^{'} a^2 \delta \varphi \bigr) \, , \nonumber
\end{eqnarray}
where $V^{'}$ denotes the derivative of $V$ with respect to $\varphi$.
These equations can be combined to give the following second order
differential equation for the relativistic potential $\phi$:
\begin{equation} \label{finaleom}
\phi^{''} + 2 \left( {\cal H} - 
{{\varphi^{''}_0} \over {\varphi^{'}_0}} \right) \phi^{'} - \nabla^2 \phi
+ 2 \left( {\cal H}^{'} - 
{\cal H}{{\varphi^{''}_0} \over {\varphi^{'}_0}} \right) \phi \, = \, 0 \, .
\end{equation}
This is the final result for the classical evolution of
cosmological fluctuations \footnote{Note that we have implicitly assumed
that the background matter field is slowly rolling, as it does in
slow-roll inflation. In the case that it is time-independent, then
the leading metric fluctuations are quadratic in the matter inhomogeneities,
as discussed in \cite{DurSak}.}. 

There are similarities between the above equation of motion for the
relativistic perturbations and
the equation (\ref{Newtoneq}) obtained in the Newtonian theory.
The final term in (\ref{finaleom}) is the force due to gravity leading
to the instability, the second to last term is the pressure force
leading to oscillations (relativistic since we are considering matter
to be a relativistic field), and the second term is the Hubble friction
term. For each wavenumber there are two fundamental solutions. On
small scales ($k > H$), the solutions correspond to damped oscillations,
on large scales ($k < H$) the oscillations freeze out and the dynamics
is governed by the gravitational force competing with the Hubble friction
term. Note, in particular, how the Hubble radius naturally emerges as
the scale where the nature of the fluctuating modes changes from oscillatory
to frozen.

Considering the equation in a bit more detail, observe that if the
equation of state of the background is independent of time (which will be
the case if ${\cal H}^{'} = \varphi^{''}_0 = 0$), then in an
expanding background, the dominant mode of (\ref{finaleom}) is constant,
and the sub-dominant mode decays. If the equation of state is not constant,
then the dominant mode is not constant in time. Specifically, at the
end of inflation ${\cal H}^{'} < 0$, and this leads to a growth of 
$\phi$ (see below).

To study the quantitative implications of the equation of motion
(\ref{finaleom}), it is convenient to introduce \cite{BST,BK}
the variable $\zeta$ (which, up to correction term of the order
$\nabla^2 \phi$ which is unimportant for large-scale fluctuations,
is equal to the curvature perturbation ${\cal R}$ in comoving gauge
\cite{Lyth}) by
\begin{equation} \label{zetaeq}
\zeta \, \equiv \, \phi + {2 \over 3} 
{{\bigl(H^{-1} {\dot \phi} + \phi \bigr)} \over { 1 + w}} \, ,
\end{equation}
where
\begin{equation} \label{wvar}
w = {p \over {\rho}}
\end{equation}
characterizes the equation of state of matter. In terms of $\zeta$,
the equation of motion (\ref{finaleom}) takes on the form
\begin{equation}
{3 \over 2} {\dot \zeta} H (1 + w) \, = \, {\cal O}(\nabla^2 \phi) \, .
\end{equation}
On large scales, the right hand side of the equation is negligible,
which leads to the conclusion that large-scale cosmological fluctuations
satisfy
\begin{equation} \label{zetacons}
{\dot \zeta} (1 + w) \, = \, 0 .
\end{equation}
This implies that $\zeta$ is constant
except possibly if $1 + w = 0$ at some point in time during the cosmological 
evolution (which occurs during reheating in inflationary
cosmology if the inflaton field undergoes oscillations - see 
\cite{Fabio1} and \cite{BaVi,Fabio2} for discussions of the consequences
in single and double field inflationary models, respectively). 
In single matter field models it is indeed possible
to show that ${\dot \zeta} = 0$ on super-Hubble scales independent
of assumptions on the equation of state \cite{Weinberg2,Zhang}.
This ``conservation law'' makes it easy to relate
initial fluctuations to final fluctuations in inflationary cosmology,
as will be illustrated in the following.

%\subsection{Application to Inflationary Cosmology}
%\label{Sec1}

Consider an application to inflationary cosmology.
We return to the space-time sketch of the evolution of
fluctuations - see Figure (\ref{fig:1}) - and use the
conservation law (\ref{zetacons}) - in the form
$\zeta = {\rm const}$ on large scales - to relate the amplitude
of $\phi$ at initial Hubble radius crossing during the inflationary
phase (at $t = t_i(k)$) with the amplitude at final Hubble radius
crossing at late times (at $t = t_f(k)$). Since both at early
times and at late times ${\dot \phi} = 0$ on super-Hubble scales,
as the equation of state is not changing, (\ref{zetacons}) and (\ref{zetaeq})
lead to
\begin{equation} \label{inflcons}
\phi(t_f(k)) \, \simeq \, {{(1 + w)(t_f(k))} \over {(1 + w)(t_i(k))}}
\phi(t_i(k)) \, .
\end{equation} 
If the initial values of the perturbations are known, then the
above equation allows us to evaluate the fluctuation amplitude 
at the time $t_f(k)$ of re-entry into the Hubble radius. 

The time-time perturbed Einstein equation (the first equation
of (\ref{perteom1})) relates the value of $\phi$ at the initial
Hubble radius crossing to the amplitude of the fractional energy
density fluctuations at that time. This, together with the fact that
the amplitude of the scalar matter field quantum vacuum fluctuations
is of the order $H$, yields
\begin{equation} \label{phiin}
\phi(t_i(k)) \, \sim \, H {{V^{'}} \over V}(t_i(k)) \, .
\end{equation}
In the late time radiation dominated phase, $w = 1/3$,
whereas during slow-roll inflation
\begin{equation} \label{win}
1 + w(t_i(k)) \, \simeq \, {{{\dot \varphi_0}^2} \over V}(t_i(k)) \, .
\end{equation}
Making, in addition, use of the slow roll conditions satisfied
during the inflationary period
\begin{eqnarray} \label{srcond}
3 H {\dot \varphi_0} \, &\simeq& \, - V^{'}  \nonumber \\
H^2 \, &\simeq& \, {{8 \pi G} \over 3} V \, ,
\end{eqnarray}
we arrive at the final result
\begin{equation} \label{final}
\phi(t_f(k)) \, \sim \, {{V^{3/2}} \over {V^{'} m_{pl}^3}}(t_i(k)) \, ,
\end{equation}
which gives the position space amplitude of cosmological
fluctuations on a scale labelled by the comoving wavenumber $k$
at the time when the scale re-enters the Hubble radius at
late times, a result first obtained in the case of the
Starobinsky model \cite{Starob} of inflation in \cite{ChibMukh},
and later in the context of scalar field-driven inflation
in \cite{GuthPi,Starob4,Hawking,BST}.

In the case of slow roll inflation, the right hand side of
(\ref{final}) is, to a first approximation, independent of $k$,
and hence the resulting spectrum of fluctuations is nearly
scale-invariant.
 
\subsection{Quantum Theory of Cosmological Fluctuations}

%\subsection{Overview}

In many models of the very early Universe, in particular
in inflationary cosmology, primordial inhomogeneities emerge
from quantum vacuum fluctuations on microscopic scales (wavelengths
smaller than the Hubble radius). The wavelength is then stretched
relative to the Hubble radius, becomes larger than the Hubble
radius at some time and the perturbation 
then propagates on super-Hubble scales until
re-entering at late cosmological times. In the context of a Universe
with a de Sitter phase, the quantum origin of cosmological fluctuations
was first discussed in \cite{ChibMukh}. In an inflationary universe,
it is easy to justify focusing attention on the quantum fluctuations:
any classical fluctuations present at the beginning of inflation
are red-shifted during the period of inflation, and will thus be
irrelevant to scales probed in observations today. On small scales,
a quantum vacuum remains.

To understand the role
of the Hubble radius, consider the equation of a free scalar matter field
$\varphi$ on an unperturbed expanding background:
\begin{equation}
\ddot{\varphi} + 3 H \dot{\varphi} - {{\nabla^2} \over {a^2}} \varphi
\, = \, 0 \, .
\end{equation}
The second term on the left hand side of this equation leads to damping
of $\varphi$ with a characteristic decay rate given by $H$. As a
consequence, in the absence of the spatial gradient term, $\dot{\varphi}$
would be of the order of magnitude $H \varphi$. Thus, comparing the
second and the third terms on the left hand side, we immediately
see that the microscopic (spatial gradient) term dominates on length
scales smaller than the Hubble radius, leading to oscillatory motion,
whereas this term is negligible on scales larger than the Hubble radius,
and the evolution of $\varphi$ is determined primarily by gravity. Note
that in general cosmological models the Hubble radius is much smaller than the
horizon (the forward light cone calculated from the initial time). In
an inflationary universe, the horizon is larger by a factor of 
${\rm exp}(N)$, where $N$ is the number of e-foldings of inflation. It is very important to realize this difference, a
difference which is obscured in most articles on cosmology in which the
term ``horizon'' is used when ``Hubble radius'' is meant. Note, in
particular, that the homogeneous inflaton field contains causal information
on super-Hubble but sub-horizon scales. Hence, it is completely consistent
with causality \cite{Fabio1}
to have a microphysical process related to the background
scalar matter field lead to exponential amplification of the amplitude
of fluctuations during reheating on such scales, as it does in models
in which entropy perturbations are present and not suppressed during
inflation \cite{BaVi,Fabio2}. 

To understand the generation and evolution of fluctuations in current
models of the very early Universe, we need both Quantum Mechanics
and General Relativity, i.e. quantum gravity. At first sight, we
are thus faced with an intractable problem, since the theory of quantum
gravity is not yet established. We are saved by the fact that today
on large cosmological scales the fractional amplitude of the fluctuations
is smaller than 1. Since gravity is a purely attractive force, the
fluctuations had to have been - at least in the context of an eternally
expanding background cosmology - very small in the early Universe. Thus,
a linearized analysis of the fluctuations (about a classical
cosmological background) is self-consistent.

From the classical theory of cosmological perturbations discussed in the
previous subsection, it follows that the analysis of scalar metric 
inhomogeneities
can be reduced - after extracting gauge artifacts -
to the study of the evolution of a single fluctuating
variable. Thus, the quantum theory of cosmological
perturbations must be reducible to the quantum theory of a single
free scalar field which we will denote by $v$. 
Since the background in which this scalar field
evolves is time-dependent, the mass of $v$ will be time-dependent. The
time-dependence of the mass will lead to quantum particle production
over time if we start the evolution in the vacuum state for $v$. As
we will see, this quantum particle production corresponds to the
development and growth of the cosmological fluctuations. 
The quantum theory of cosmological fluctuations provides a consistent
framework to study both the generation and the evolution of metric
perturbations.
 
In order to obtain the action for linearized cosmological
perturbations, we expand the action to quadratic order in
the fluctuating degrees of freedom. The linear terms cancel
because the background is taken to satisfy the background
equations of motion.

We begin with the Einstein-Hilbert action for gravity and the
action of a scalar matter field 
\begin{equation} \label{action}
S \, = \,  \int d^4x \sqrt{-g} \bigl[ - {1 \over {16 \pi G}} R
+ {1 \over 2} \partial_{\mu} \varphi \partial^{\mu} \varphi - V(\varphi)
\bigr] \, ,
\end{equation}
where $R$ is the Ricci curvature scalar.

The simplest way to proceed is to work in 
longitudinal gauge.
The next step is to reduce the number of degrees of freedom.
Since for scalar field matter there are no anisotropic stresses
to linear order, the off-diagonal
spatial Einstein equations force $\psi = \phi$. 
The two remaining fluctuating variables
$\phi$ and $\varphi$ are linked by the Einstein constraint
equations since there cannot be matter fluctuations without induced
metric fluctuations. 

The two nontrivial tasks of the lengthy \cite{MFB} computation 
of the quadratic piece of the action is to find
out what combination of $\varphi$ and $\phi$ yields the variable $v$
in terms of which the action has canonical kinetic term, and what the form
of the time-dependent mass is. In the context of
scalar field matter, the quantum theory of cosmological
fluctuations was developed by Mukhanov \cite{Mukh2,Mukh3} and
Sasaki \cite{Sasaki}. The result is the following
form of the action quadratic in the
perturbations:
\begin{equation} \label{pertact}
S^{(2)} \, = \, {1 \over 2} \int d^4x \bigl[v'^2 - v_{,i} v_{,i} + 
{{z''} \over z} v^2 \bigr] \, ,
\end{equation}
where the canonical variable $v$ (the ``Sasaki-Mukhanov variable'' introduced
in \cite{Mukh3} - see also \cite{Lukash}) is given by
\begin{equation} \label{Mukhvar}
v \, = \, a \bigl[ \delta \varphi + {{\varphi_0^{'}} \over {\cal H}} \phi
\bigr] \, 
\end{equation}
and 
\begin{equation} \label{zvar}
z \, = \, {{a \varphi_0^{'}} \over {\cal H}} \, .
\end{equation}

As long as
the equation of state does not change over time
${\cal H}$ and $\varphi_0^{'}$ are proportional and hence
\begin{equation} \label{zaprop}
z(\eta) \, \sim \, a(\eta) \, .
\end{equation}
Note that the variable $v$ is related to the curvature
perturbation ${\cal R}$ in comoving coordinates introduced
in \cite{Lyth} and closely related to the variable $\zeta$ used
in \cite{BST,BK}:
\begin{equation} \label{Rvar}
v \, = \, z {\cal R} \, .
\end{equation}

The equation of motion which follows from the action (\ref{pertact}) is
(in momentum space)
\begin{equation} \label{pertEOM2}
v_k^{''} + k^2 v_k - {{z^{''}} \over z} v_k \, = \, 0 \, ,
\end{equation}
where $v_k$ is the k'th Fourier mode of $v$. As a consequence of
(\ref{zaprop}), the tachyonic mass term in the above equation is given
by the Hubble scale
\begin{equation}
k_H^2 \, \equiv \, {{z^{''}} \over z} \, \simeq \, H^2 \, .
\end{equation}
Thus, it immediately follows from (\ref{pertEOM2}) that on small
length scales, i.e. for
$k > k_H$, the solutions for $v_k$ are constant amplitude oscillations . 
These oscillations freeze out at Hubble radius crossing,
i.e. when $k = k_H$. On longer scales ($k \ll k_H$), the solutions
for $v_k$ increase as $z$:
\begin{equation} \label{squeezing}
v_k \, \sim \, z \,\, , \,\,\,   k \ll k_H \, .
\end{equation}
The state of the fluctuations becomes a squeezed quantum state.

Given the action (\ref{pertact}), the quantization of the cosmological
perturbations can be performed by canonical quantization (in the same
way that a scalar matter field on a fixed cosmological background
is quantized \cite{BD}). 

The final step in the quantum theory of cosmological perturbations
is to specify an initial state. Since in inflationary cosmology
all pre-existing classical fluctuations are red-shifted by the
accelerated expansion of space, one usually assumes (we will
return to a criticism of this point when discussing the
trans-Planckian problem of inflationary cosmology) that the field
$v$ starts out at the initial time $t_i$ mode by mode in its vacuum
state. Two questions immediately emerge: what is the initial time $t_i$,
and which of the many possible vacuum states should be chosen. It is
usually assumed that since the fluctuations only oscillate on sub-Hubble
scales, the choice of the initial time is not important, as long
as it is earlier than the time when scales of cosmological interest
today cross the Hubble radius during the inflationary phase. The
state is usually taken to be the Bunch-Davies vacuum (see e.g. \cite{BD}),
since this state is empty of particles at $t_i$ in the coordinate frame
determined by the FRW coordinates (see e.g. \cite{RB84} for a
discussion of this point), and since the Bunch-Davies state is
a local attractor in the space of initial states in an expanding
background (see e.g. \cite{BHill}). Thus, we choose the initial
conditions
\begin{eqnarray} \label{incond}
v_k(\eta_i) \, = \, {1 \over {\sqrt{2 \omega_k}}} \\
v_k^{'}(\eta_i) \, = \, {{\sqrt{\omega_k}} \over {\sqrt{2}}} \, \, \nonumber
\end{eqnarray}
where here $\omega_k = k$, and $\eta_i$ is the conformal time corresponding
to the physical time $t_i$.

Let us briefly summarize the quantum theory of cosmological perturbations.
In the linearized theory, fluctuations are set up at some initial
time $t_i$ mode by mode in their vacuum state. While the wavelength
is smaller than the Hubble radius, the state undergoes quantum
vacuum fluctuations. The accelerated expansion of the
background redshifts the length scale beyond the Hubble radius. The 
fluctuations freeze out when the length scale is equal to the Hubble
radius. On larger scales, the amplitude of $v_k$ increases as the
scale factor. This corresponds to the squeezing of the quantum state present
at Hubble radius crossing (in terms of classical general relativity, it
is self-gravity which leads to this growth of fluctuations). As
discussed e.g. in \cite{PolStar}, the squeezing of the quantum vacuum state
sets up the classical correlations in the wave function of the
fluctuations which are an essential ingredient in the classicalization of
the perturbations. Squeezing also leads to the phase coherence of
fluctuations on all scales, which in turn is responsible for producing
the oscillations in the CMB angular power spectrum, as realized long
before the advent of inflationary cosmology in \cite{SZ,Peebles}.

We end this subsection with a calculation of the spectrum
of curvature fluctuations in inflationary cosmology. 

We need to compute the power spectrum ${\cal P}_{\cal R}(k)$ 
of the curvature fluctuation ${\cal R}$ defined in (\ref{Rvar}).
The idea in calculating the power spectrum at a late time $t$ is
to first relate the power spectrum via the growth rate (\ref{squeezing})
of $v$ on super-Hubble scales to the power spectrum at the time $t_H(k)$
of Hubble radius crossing, and to then use the constancy of the amplitude
of $v$ on sub-Hubble scales to relate it to the initial conditions
(\ref{incond}). Thus
\begin{eqnarray} \label{finalspec1}
2 \pi^2 {\cal P}_{\cal R}(k, t) \, \equiv  \, k^3 |{\cal R}_k|^2(t) \, 
&=& \, k^3 z^{-2}(t) |v_k(t)|^2 \\
&=& \, k^3 z^{-2}(t) \bigl( {{z(t)} \over {z(t_H(k))}} \bigr)^2
|v_k(t_H(k))|^2 \nonumber \\
&=& \, k^3 z^{-2}(t_H(k)) |v_k(t_H(k))|^2 \nonumber \\
&\sim& \, k^3 a^{-2}(t_H(k)) |v_k(t_i)|^2 \, , \nonumber
\end{eqnarray}
where in the final step we have used (\ref{zaprop}) and the
constancy of the amplitude of $v$ on sub-Hubble scales. Making use
of the condition
$a^{-1}(t_H(k)) k \, = \, H$
for Hubble radius crossing, and of the
initial conditions (\ref{incond}), we immediately see that
\begin{equation} \label{finalspec2}
{\cal P}_{\cal R}(k, t) \, \sim \, k^3 k^{-2} k^{-1} H^2 \, ,
\end{equation}
and that thus a scale invariant power spectrum with amplitude
proportional to $H^2$ results.

The quantization of gravitational waves parallels the
quantization of scalar metric fluctuations, but is
more simple because there are no gauge ambiguities. The 
starting point is the action (\ref{action}), into which
we insert the metric which corresponds to a 
classical cosmological background plus tensor metric
fluctuations:
\begin{equation}
ds^2 \, = \, a^2(\eta) \bigl[ d\eta^2 - (\delta_{ij} + h_{ij}) dx^i dx^j 
\bigr]\, ,
\end{equation}
where the second rank tensor $h_{ij}(\eta, {\bf x})$ represents the
gravitational waves, and in turn can be decomposed as
\begin{equation}
h_{ij}(\eta, {\bf x}) \, = \, h_{+}(\eta, {\bf x}) e^+_{ij}
+ h_{x}(\eta, {\bf x}) e^x_{ij}
\end{equation}
into the two polarization states. Here, $e^{+}_{ij}$ and $e^{x}_{ij}$ are
two fixed polarization tensors, and $h_{+}$ and $h_{x}$ are the two 
coefficient functions.

To quadratic order in the fluctuating fields, the action consists of
separate terms involving $h_{+}$ and $h_{x}$. Each term is of the form
\begin{equation} \label{actgrav}
S^{(2)} \, = \, \int d^4x {{a^2} \over 2} \bigl[ h'^2 - (\nabla h)^2 \bigr] 
\, ,
\end{equation}
leading to the equation of motion
\begin{equation}
h_k^{''} + 2 {{a'} \over a} h_k^{'} + k^2 h_k \, = \, 0 \, .
\end{equation}
The variable in terms of which the action (\ref{actgrav}) has canonical
kinetic term is
\begin{equation} \label{murel}
\mu_k \, \equiv \, a h_k \, ,
\end{equation}
and its equation of motion is
\begin{equation}
\mu_k^{''} + \bigl( k^2 - {{a''} \over a} \bigr) \mu_k \, = \, 0 \, .
\end{equation}
This equation is very similar to the corresponding equation (\ref{pertEOM2}) 
for scalar gravitational inhomogeneities, except that in the mass term
the scale factor $a(\eta)$ replaces $z(\eta)$, which leads to a
very different evolution of scalar and tensor modes during the reheating
phase in inflationary cosmology during which the equation of state of the
background matter changes dramatically.
 
Based on the above discussion we have the following theory for the
generation and evolution of gravitational waves in an accelerating
Universe (first developed by Grishchuk \cite{Grishchuk}): 
waves exist as quantum vacuum fluctuations at the initial time
on all scales. They oscillate until the length scale crosses the Hubble
radius. At that point, the oscillations freeze out and the quantum state
of gravitational waves begins to be squeezed in the sense that
\begin{equation}
\mu_k(\eta) \, \sim \, a(\eta) \, ,
\end{equation}
which, from (\ref{murel}) corresponds to constant amplitude of $h_k$.
The squeezing of the vacuum state leads to the emergence of classical
properties of this state, as in the case of scalar metric fluctuations.

\section{Towards a New Paradigm: String Gas Cosmology}

In spite of the phenomenological successes of inflationary cosmology
in addressing some key problems of standard cosmology and in providing
a predictive theory of structure formation, current models of inflation
are faced with some key conceptual problems which motivate the search
for a new early universe paradigm based on new fundamental physics,
string theory being the most promising candidate.
Below, we first list some of the key problems for inflation, and then
discuss a toy model designed to explore possible cosmological consequences
of some of the key new features of string theory.

\subsection{Conceptual Problems of Scalar Field-Driven Inflation}

\centerline{\it Nature of the Inflaton}

In the context of General Relativity as the theory of space-time,
matter with an equation of state $p \, \simeq \, - \rho$ is
required in order to obtain almost exponential expansion of space.
If we describe matter in terms of fields with canonical kinetic
terms, a scalar field is required since in the context of usual
renormalizable field theories it is only for scalar fields 
that a potential energy
function in the Lagrangian is allowed, and of all energy
terms only the potential energy can yield the 
required equation of state. 

In order for scalar fields to generate a period of cosmological
inflation, the potential energy needs to dominate over the kinetic
and spatial gradient energies. It is generally assumed that spatial
gradient terms can be neglected. This is, however, not true in general.
Next, assuming a homogeneous field configuration, we must ensure that the
potential energy dominates over the kinetic energy. This leads to
the first ``slow-roll" condition. Requiring the period of inflation
to last sufficiently long leads to a second slow-roll condition, namely
that the $\ddot \varphi$ term in the Klein-Gordon equation for the
inflaton $\varphi$ be negligible. Scalar fields charged with respect
to the Standard Model symmetry groups do not satisfy the slow-roll
conditions. 

Assuming that both slow-roll conditions hold, one
obtains a ``slow-roll trajectory" in the phase space of
homogeneous $\varphi$ configurations. In large-field inflation models
such as ``chaotic inflation" \cite{chaotic}
and ``hybrid inflation" \cite{hybrid}, the slow-roll
trajectory is a local attractor in initial condition space \cite{Kung}
(even when linearized metric perturbations are taken into account
\cite{Feldman}), whereas this is not the case \cite{Goldwirth}
in small-field models such as ``new inflation" \cite{new}. As
shown in \cite{GhazalScott}, this leads to problems for some
models of inflation which have recently been proposed in the context
of string theory. To address this problem, it has been proposed
that inflation may be future-eternal \cite{eternal} and that it is
hence sufficient that there be some configurations in initial condition
space which give rise to inflation within one Hubble patch, inflation
being then self-sustaining into the future. However, one must still
ensure that slow-roll inflation can locally be satisfied.

Many models of particle physics beyond the Standard Model contain a
plethora of new scalar fields. One of the most conservative extensions
of the Standard Model is the MSSM, the ``Minimal Supersymmetric Standard
Model". According to a recent study, among the many scalar fields in this
model, only a hand-full can be candidates for a slow-roll inflaton,
and even then very special initial conditions are required \cite{MSSM}.
The situation in supergravity and superstring-inspired field theories
may be more optimistic, but the issues are not settled (see e.g.
\cite{Cliff,Jim,Andrei} for recent reviews).

\centerline{\it Hierarchy Problem}
 
Assuming for the sake of argument that a successful model of
slow-roll inflation has been found, one must still build in
a hierarchy into the field theory model in order to obtain an
acceptable amplitude of the density fluctuations (this is
sometimes also called the ``amplitude problem"). Unless
this hierarchy is observed, the density fluctuations will
be too large and the model is observationally ruled out.
 
In a wide class of inflationary
models, obtaining the correct amplitude requires the introduction
of a hierarchy in scales, namely \cite{Adams}
\begin{equation}
{{V(\varphi)} \over {\Delta \varphi^4}}
\, \leq \, 10^{-12} \, ,
\end{equation}
where $\Delta \varphi$ is the change in the inflaton field during
one Hubble expansion time (during inflation), and $V(\varphi)$ is
the potential energy during inflation.

This problem should be contrasted with the success of topological
defect models (see e.g. \cite{ShellVil,HK,RHBrev0} for
reviews) in predicting the right oder of magnitude of
density fluctuations without introducing a new scale of physics.
The GUT scale as the scale of the symmetry breaking phase transition 
(which produces the defects) yields the correct magnitude of the
spectrum of density fluctuations \cite{CSpapers}. Topological
defects, however, cannot be the prime mechanism for the origin of
fluctuations since they do not give rise to 
coherent adiabatic fluctuations and hence fail to yield
acoustic oscillations
in the angular power spectrum of the CMB anisotropies \cite{CSanis}.

\centerline{\it Trans-Planckian Problem}

A more serious problem is the ``trans-Planckian problem" \cite{RHBrev1}.
Returning to the space-time diagram of Figure 1, we can immediately
deduce that, provided that the period of inflation lasted sufficiently
long (for GUT scale inflation the number is about 70 e-foldings),
then all scales inside of the Hubble radius today started out with a
physical wavelength smaller than the Planck scale at the beginning of
inflation. Now, the theory of cosmological perturbations is based
on Einstein's theory of General Relativity coupled to a simple
semi-classical description of matter. It is clear that these
building blocks of the theory are inapplicable on scales comparable
and smaller than the Planck scale. Thus, the key
successful prediction of inflation (the theory of the origin of
fluctuations) is based on suspect calculations since 
new physics {\it must} enter
into a correct computation of the spectrum of cosmological perturbations.
The key question is as to whether the predictions obtained using
the current theory are sensitive to the specifics of the unknown
theory which takes over on small scales.
 
One approach to study the sensitivity of the usual predictions of
inflationary cosmology to the unknown physics on trans-Planckian scales
is to study toy models of ultraviolet physics which allow explicit
calculations. The first approach which was used \cite{Jerome1,Niemeyer}
is to replace the usual linear dispersion relation for the Fourier
modes of the fluctuations by a modified dispersion relation, a
dispersion relation which is linear for physical wavenumbers smaller
than the scale of new physics, but deviates on larger scales. Such
dispersion relations were used previously to test the sensitivity
of black hole radiation on the unknown physics of the ultra-violet 
\cite{Unruh,CJ}. It was found \cite{Jerome1} that if the 
evolution of modes on
the trans-Planckian scales is non-adiabatic, then substantial
deviations of the spectrum of fluctuations from the usual results
are possible. Non-adiabatic evolution turns an initial state
minimizing the energy density into a state which is excited once
the wavelength becomes larger than the cutoff scale. Back-reaction
effects of these excitations may limit the magnitude of the
trans-Planckian effects, but - based on our recent study \cite{Jerome3} -
not to the extent initially expected \cite{Tanaka,Starob3}.

From the point of view of fundamental physics, the {\it trans-Planckian
problem} is not a problem. Rather, it yields a window of opportunity to
probe new fundamental physics in current and future observations, even
if the scale of the new fundamental physics is close to the Planck scale.
The point is that if the universe in fact underwent a period of inflation,
then trans-Planckian physics leaves an imprint on the spectrum of fluctuations.
The exponential expansion of space amplifies the wavelength of the 
perturbations
to observable scales. At the present time, it is our ignorance about
quantum gravity which prevents us from making any specific predictions. For
example, we do not understand string theory in time-dependent backgrounds 
sufficiently well to be able to at this time make any predictions for
observations. 

\centerline{\it Singularity Problem}

The next problem is the ``singularity problem". This problem, one
of the key problems of Standard Cosmology, has not been resolved in
models of scalar field-driven inflation.

As follows from the Penrose-Hawking singularity theorems of General
Relativity (see e.g. \cite{HE} for a textbook discussion), an initial
cosmological singularity is unavoidable if space-time is described in
terms of General Relativity, and if the matter
sources obey the weak energy conditions. 
Recently, the singularity theorems have been
generalized to apply to Einstein gravity coupled to scalar field
matter, i.e. to scalar field-driven inflationary cosmology \cite{Borde}.
It is shown that in this context, a past singularity at some point
in space is unavoidable. 

In the same way that the appearance of an initial singularity in Standard
Cosmology told us that Standard Cosmology cannot be the correct description of
the very early universe, the appearance of an initial singularity in
current models of inflation tell us that inflationary cosmology cannot yield
the correct description of the very, very early universe. At sufficiently
high densities, a new description will take over. In the same way
that inflationary cosmology contains late-time standard cosmology, it is
possible that the new cosmology will contain, at later times, inflationary
cosmology. However, one should keep an open mind to the possibility that
the new cosmology will connect to present observations via a route which
does not contain inflation, a possibility explored later in this
section.

\centerline{\it Breakdown of Validity of Einstein Gravity}

The Achilles heel of scalar field-driven inflationary cosmology is,
however, the use of intuition from Einstein gravity at energy scales
not far removed from the Planck and string scales, scales where
correction terms to the Einstein-Hilbert term in the gravitational
action dominate and where intuition based on applying the Einstein
equations break down (see also \cite{swamp} for arguments along
these lines). 

All approaches to quantum gravity predict correction terms in the
action which dominate at energies close to the Planck scale - in
some cases in fact even much lower. Semiclassical gravity leads to
higher curvature terms, and may (see e.g. \cite{BMS,Biswas}) lead
to bouncing cosmologies without a singularity). Loop quantum
cosmology leads to similar modifications of early universe cosmology
(see e.g. \cite{Bojowald} for a recent review). String theory,
the theory we will focus on in the following sections, has a
maximal temperature for a string gas in thermal equilibrium \cite{Hagedorn},
which may lead to an almost static phase in the early universe - the
Hagedorn phase \cite{BV}. 

Common to all of these approaches to quantum gravity corrections to
early universe cosmology is the fact that a transition from a contracting (or
quasi-static) early universe phase to the rapidly expanding radiation phase
of standard cosmology can occur {\it without} violating the usual energy
conditions for matter. In particular, it is possible (as is predicted
by the string gas cosmology model discussed below) that the universe in an
early high temperature phase is almost static. This may be a common feature
to a large class of models which resolve the cosmological singularity.

Closely related to the above is the ``cosmological constant problem" for
inflationary cosmology. We know from
observations that the large quantum vacuum energy of field theories
does not gravitate today. However, to obtain a period of inflation
one is using precisely the part of the energy-momentum tensor of the 
inflaton field
which looks like the vacuum energy. In the absence of a convincing
solution of the cosmological constant problem it is unclear whether
scalar field-driven inflation is robust, i.e. whether the
mechanism which renders the quantum vacuum energy gravitationally
inert today will not also prevent the vacuum energy from
gravitating during the period of slow-rolling of the inflaton 
field \footnote{Note that the
approach to addressing the cosmological constant problem making use
of the gravitational back-reaction of long range fluctuations
(see \cite{RHBrev4} for a summary of this approach) does not prevent
a long period of inflation in the early universe.}.

\subsection{String Gas Cosmology}

An immediate problem which arises when trying to connect string theory
with cosmology is the {\it dimensionality problem}. Superstring theory
is perturbatively consistent only in ten space-time dimensions, but we
only see three large spatial dimensions. The original approach to 
addressing this problem was to assume that the six extra dimensions are
compactified on a very small space which cannot be probed with our
available energies. However, from the point of view of cosmology,
it is quite unsatisfactory not to be able to understand why it is
precisely three dimensions which are not compactified and why the compact
dimensions are stable. Brane world cosmology \cite{branereview} provides
another approach to this problem: it assumes that we live on a
three-dimensional brane embedded in a large nine-dimensional space.
Once again, a cosmologically satisfactory theory should explain
why it is likely that we will end up exactly on a three-dimensional
brane (for some interesting work addressing this issue see
\cite{Mahbub,Mairi,Lisa}).

Finding a natural solution to the dimensionality problem is thus one
of the key challenges for superstring cosmology. This challenge has
various aspects. First, there must be a mechanism which singles out
three dimensions as the number of spatial dimensions we live in.
Second, the moduli fields which describe the volume and the shape of
the unobserved dimensions must be stabilized (any strong time-dependence
of these fields would lead to serious phenomenological constraints).
This is the {\it moduli problem} for superstring cosmology. As
mentioned above, solving the {\it singularity problem} is another of
the main challenges. These are the three problems which {\it string
gas cosmology} \cite{BV,TV,ABE} explicitly addresses at the present
level of development \footnote{See also \cite{Rama} for
an alternative approach within string theory to explaining why
only three spatial dimensions are large.}.

In the absence of a non-perturbative formulation of string theory,
the approach to string cosmology which we have suggested,
{\it string gas cosmology} \cite{BV,TV,ABE} (see also \cite{Perlt}
for early work, and \cite{BattWat,RHBrev5,RHBrev6,RHBrev7} for reviews),
is to focus on symmetries and degrees of freedom which are new to
string theory (compared to point particle theories) and which will
be part of a non-perturbative string theory, and to use
them to develop a new cosmology. The symmetry we make use of is
{\it T-duality}, and the new degrees of freedom are 
{\it string winding modes} and {\it string oscillatory modes}.

We take all spatial directions to be toroidal, with
$R$ denoting the radius of the torus. Strings have three types
of states: {\it momentum modes} which represent the center
of mass motion of the string, {\it oscillatory modes} which
represent the fluctuations of the strings, and {\it winding
modes} counting the number of times a string wraps the torus.
Both oscillatory and winding states are special to strings.
Point particle theories do not contain these modes. 

The energy of an oscillatory mode is independent of $R$, momentum
mode energies are quantized in units of $1/R$, i.e.
\be
E_n \, = \, n {1 \over R} \, ,
\ee
whereas the winding mode energies are quantized in units of $R$, i.e.
\be
E_m \, = \, m R \, ,
\ee
where both $n$ and $m$ are integers. The energy of oscillatory modes
does not depend on $R$. 

The T-duality symmetry is the invariance of the spectrum of string
states under the change
\be \label{Tdual}
R \, \rightarrow \, 1/R
\ee
in the radius of the torus (in units of the string length $l_s$).
Under such a change, the energy spectrum of string states is
not modified if winding and momentum quantum numbers are interchanged
\be
(n, m) \, \rightarrow \, (m, n) \, .
\ee
The string vertex operators are consistent with this symmetry, and
thus T-duality is a symmetry of perturbative string theory. Postulating
that T-duality extends to non-perturbative string theory leads
\cite{Pol} to the need of adding D-branes to the list of fundamental
objects in string theory. With this addition, T-duality is expected
to be a symmetry of non-perturbative string theory.
Specifically, T-duality will take a spectrum of stable Type IIA branes
and map it into a corresponding spectrum of stable Type IIB branes
with identical masses \cite{Boehm}.

Since the number of string oscillatory modes increases exponentially
as the string mode energy increases, there is a maximal temperature
of a gas of strings in thermal equilibrium, the {\it Hagedorn
temperature} $T_H$ \cite{Hagedorn}. If we imagine taking a box of strings
and compressing it, the temperature will never exceed $T_H$. In fact,
as the radius $R$ decreases below the string radius, the temperature
will start to decrease, obeying the duality relation \cite{BV}
\be
T(R) \, = \, T(1/R) \, .
\ee
This argument shows that string theory has the potential of
taming singularities in physical observables. 
Figure 4 provides a sketch of how the temperature $T$ changes 
as a function of $R$.
\begin{figure}
\centering
%\centerline{\epsfxsize=3in\epsfbox{canc1.eps}}
\includegraphics[height=6cm]{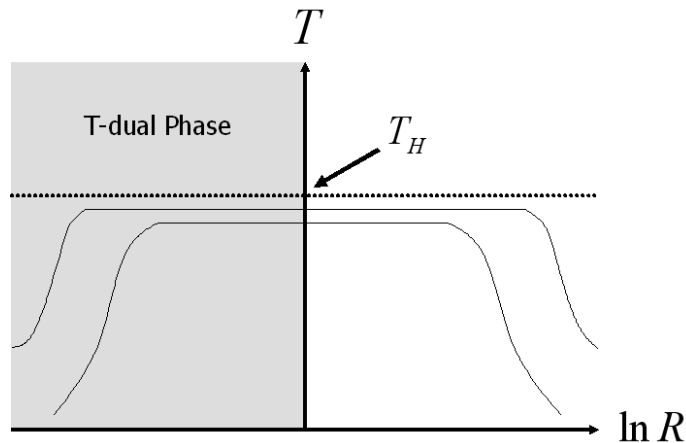}
\caption{Sketch (based on the analysis of \cite{BV}
of the evolution of temperature $T$ as a function
of the radius $R$ of space of a gas of strings in thermal
equilibrium. The top curve is characterized by an entropy
higher than the bottom curve, and leads to a longer
region of Hagedorn behaviour.}
\label{fig:5}       
\end{figure}

If we imagine that there is a dynamical principle that tells us how
$R$ evolves in time, then Figure 2 can be interpreted as depicting
how the temperature changes as a function of time. If $R$ is a monotonic
function of time, then two interesting possibilities for cosmology
emerge. If ${\rm ln}R$ decreases to zero at some fixed time (which without
loss of generality we can call $t = 0$),
and continues to decrease, we obtain a temperature profile which is
symmetric with respect to $t = 0$ and which (since small $R$ is physically
equivalent to large $R$) represents a bouncing cosmology 
(see \cite{Biswas2}
for a concrete recent realization of this scenario). If, on the other hand,
it takes an inifinite amount of time to reach $R = 0$, an {\it emergent
universe} scenario \cite{gellis} is realized.

It is important to realize that in both of the cosmological
scnearios which, as argued above, seem to follow from string theory symmetry
considerations alone, a large energy density does {\it not} lead
to rapid expansion in the Hagedorn phase, in spite of the fact that
the matter sources we are considering (namely a gas of strings) obey all
of the usual energy conditions discussed e.g. in \cite{HE}). These
considerations are telling us that intuition drawn from Einstein gravity
will give us a completely incorrect picture of the early universe. 
 
Any physical theory requires both a specification of the equations of
motion and of the initial conditions.  We assume that
the universe starts out small and hot. For simplicity, we take
space to be toroidal, with radii in all spatial directions given by
the string scale. We assume that the initial energy density 
is very high, with an effective temperature which is close
to the Hagedorn temperature, the maximal temperature
of perturbative string theory. 

In this context, it was argued \cite{BV} that in order for spatial 
sections to become
large, the winding modes need to decay. This decay, at least
on a background with stable one cycles such as a torus, is only
possible if two winding modes meet and annihilate. Since string
world sheets have measure zero probability for intersecting in more
than four space-time dimensions, winding modes can annihilate only
in three spatial dimensions (see, however, the recent
caveats to this conclusion based on the work of \cite{Kabat3,Danos}). 
Thus, only three spatial dimensions
can become large, hence explaining the observed dimensionality of
space-time. As was shown later \cite{ABE}, adding branes to
the system does not change these conclusions since at later
times the strings dominate the cosmological dynamics.
Note that in the three dimensions which are becoming large there
is a natural mechanism of isotropization as long as some winding
modes persist \cite{Watson1}.

Some of the above heuristic arguments can be
put on a more firm mathematical basis, albeit in the context
of a toy model, a model consisting of a classical
background coupled to a gas of strings. From the point of
view of rigorous string theory, this separation between
classical background and stringy matter is not satisfactory
when dealing with very early times when the typical length
scale might be the string scale..
However, in the absence of a non-perturbative formulation of
string theory, at the present time we are forced to make this
separation. Note that this separation between classical
background geometry and string matter is common to all current
approaches to string cosmology.

The background is described by dilaton gravity. The
dilaton must be included since it arises in string theory
at the same level as the graviton, and also (in the context
of string gas cosmology) because it and not Einstein gravity  
is consistent with the T-duality symmetry. 
Note, however, that the background dynamics inevitably drives
the system into a parameter region where the dilaton is
strongly coupled and hence beyond the region of validity
of the approximations made.
 
The action for dilaton gravity  
coupled to a matter action $S_m$ is
\be
S \, = \, {1 \over {2 \kappa^2}} \int d^{10}x \sqrt{-g} e^{-2 \phi}
\bigl[R + 4 \partial^{\mu} \phi \partial_{\mu} \phi \bigr] + S_m \, ,
\ee
where $g$ is the determinant of the metric, 
$R$ is the Ricci scalar,
$\phi$ is the dilaton, and
$\kappa$ is the reduced gravitational constant in ten dimensions.
The metric appearing in the above action is the metric in the
string frame. 

For a homogeneous and isotropic metric
\be
ds^2 \, = \, dt^2 - a(t)^2 d{\bf x}^2 \, ,
\ee
the resulting equations of motion in the string frame are
\cite{TV} (see also \cite{Ven})
\bea
-d {\dot \lambda}^2 + {\dot \varphi}^2 \, &=& \, e^{\varphi} E 
\label{E1} \\
{\ddot \lambda} - {\dot \varphi} {\dot \lambda} \, &=& \,
{1 \over 2} e^{\varphi} P \label{E2} \\
{\ddot \varphi} - d {\dot \lambda}^2 \, &=& \, {1 \over 2} e^{\varphi} E \, ,
\label{E3}
\eea
where $E$ and $P$ denote the total energy and pressure, respectively,
$d$ is the number of spatial dimensions, and we have introduced the
logarithm of the scale factor 
\be
\lambda(t) \, = \, {\rm log} (a(t))
\ee
and the rescaled dilaton
\be
\varphi \, = \, 2 \phi - d \lambda \, .
\ee

The second of these equations indicates
that a gas of strings containing both stable winding and
momentum modes will lead to the stabilization of the
radius of the torus: windings prevent expansion, momenta
prevent the contraction. The right hand side of the equation
can be interpreted as resulting from a confining potential for
the scale factor. One of the key issues when dealing with theories
with extra dimensions is the question of how the size and
shape moduli of the extra-dimensional spaces are stabilized.
String gas cosmology provides a simple and string-specific
mechanism to stabilize most of these moduli. This topic
will not be reviewed here (see \cite{RHBrev6,RHBrev7}
for recent reviews). The outstanding issue is how to
stabilize the dilaton.

Note that the dilaton is evolving at the time when
the radius of the torus is at the minimum of its potential.
For the branch of solutions we are considering,
the dilaton is increasing as we go into the past. At
some point, therefore, it becomes greater than zero. At
this point, we enter the region of strong coupling. As
already discussed in \cite{Riotto}, a different dynamical
framework is required to analyze this phase. In
particular, the fundamental
strings are no longer the lightest degrees of freedom. We
will call this phase the ``strongly coupled Hagedorn phase"
\cite{Betal} for which we lack an analytical description. Since the
energy density in this phase is of the string scale, the
background equations should also be very different from
the dilaton gravity equations used above. In the following,
we assume that the dilaton is
frozen in the strongly coupled Hagedorn phase.
This could be a consequence of S-duality (see e.g. \cite{Kaya06}).

\subsection{String Gas Cosmology and Structure Formation}

The following are key aspects of the string gas
cosmology background which emerge from the previous
discussion. First, in thermal equilibrium at the 
string scale ($R \simeq l_s$), the
self-dual radius, the number of winding and momentum modes are 
equal. Since winding and momentum modes give an opposite 
contribution to the pressure, the pressure of the string gas in 
thermal equilibrium at the self-dual radius will vanish. From  
the dilaton gravity equations of motion (\ref{E1} - \ref{E3})
it then follows that a static phase 
$\lambda = 0$ will be a fixed point of the dynamical system. This 
phase is the Hagedorn phase. 
 
On the other hand, for large values of $R$ in thermal equilibrium 
the energy will be exclusively in momentum modes. These act as usual 
radiation. Inserting the radiative equation of state into the above 
equations (\ref{E1} - \ref{E3}) it follows that the source in 
the dilaton equation of motion vanishes and the dilaton approaches 
a constant as a consequence of the Hubble damping term in its
equation of motion. Consequently, the 
scale factor expands as in the usual radiation-dominated universe.
The transition between the Hagedorn phase and the radiation-dominated 
phase with fixed dilaton is achieved via the annihilation of winding 
modes, as studied in detail in \cite{BEK}. 
The main point is that, 
starting in a Hagedorn phase, there will be a smooth transition to 
the radiation-dominated phase of standard cosmology with fixed dilaton. 

Our new cosmological background is obtained by following our
currently observed universe into the past according to the
string gas cosmology equations. The radiation phase of standard
cosmology is unchanged. In particular, the dilaton is fixed
in this phase \footnote{The dilaton comes to rest, but it is
not pinned to a particular value by a potential. Thus, in order
to obtain consistency with late time cosmology, an additional
mechanism operative at late times which fixes the dilaton is
required.}. However, as the temperature of the radiation
bath approaches the Hagedorn temperature, the equation of state
of string gas matter changes. The equation of state parameter
$w = P / E$ decreases towards a pressureless state and the
string frame metric becomes static. Note that, in order
for the present size of the universe to be larger than our current
Hubble radius, the size of the spatial sections in the
Hagedorn phase must be at least $1 {\rm mm}$
\footnote{How to obtain this initial size starting from string-scale
initial conditions constitutes the {\it entropy problem} of
our scenario. A possible solution making use of an initial phase of
bulk dynamics is given in \cite{Natalia}.}. We have entered
the Hagedorn phase. 

As we go back
in time in the Hagedorn phase, the dilaton increases. At the time
$t_c$ when the dilaton equals zero, a second transition occurs, the
transition to a ``strongly coupled Hagedorn phase'' (using the
terminology introduced in \cite{Betal}). We take the dilaton to be
fixed in this phase. In this case, the strongly coupled Hagedorn
phase may have a duration which is very long compared to the 
Einstein frame Hubble time immediately following $t_R$. 
It is in this cosmological background that we will study the 
generation of fluctuations.  We will denote
the time when the Einstein frame Hubble radius reaches a minimum
by $t_R$, to evoque the analogy
with the time of reheating in inflationary cosmology. The end of the
Hagedorn phase (the time
when the radiation phase of standard cosmology begins) is slightly
later. 

It is instructive to compare the background evolution of string gas 
cosmology with the background of inflationary cosmology. 
Figure 5 is a sketch of 
the space-time evolution in string gas cosmology. For times $t < t_R$, 
we are in the static Hagedorn phase and the Hubble radius is
infinite. For $t > t_R$, the Einstein frame 
Hubble radius is expanding as in standard cosmology. To understand 
why string gas cosmology can lead to a causal mechanism of structure 
formation, we must compare the evolution of the physical 
wavelength corresponding to a fixed comoving scale  
with that of the Einstein frame Hubble radius $H^{-1}(t)$. 
Recall that the Einstein frame Hubble radius separates 
scales on which fluctuations oscillate (wavelengths smaller than 
the Hubble radius) from wavelengths on which the fluctuations are frozen 
in and cannot be effected by microphysics. Causal microphysical 
processes can generate fluctuations only on sub-Hubble scales 
\footnote{Matter which couples minimally to gravity in the string frame
is prevented from oscillating on scales larger than the Einstein frame
Hubble radius by the dilaton friction term in its equation of motion.}. 
The key point is that for $t < t_i(k)$, the fluctuation mode $k$ is inside
the Hubble radius, and thus a causal generation mechanism for fluctuations
is possible.
 
\begin{figure} 
\includegraphics[height=10cm]{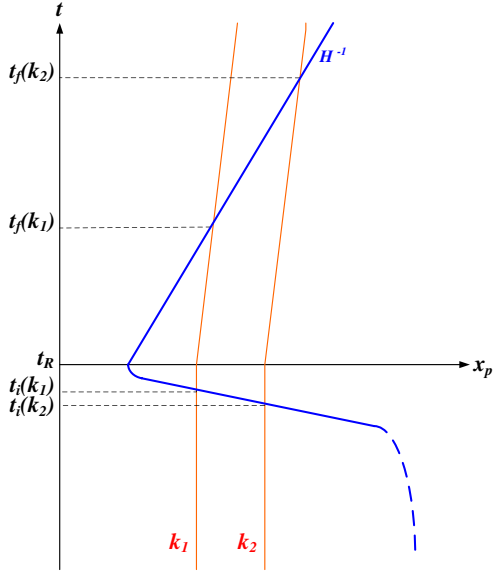} 
\caption{Space-time diagram (sketch) showing the evolution of fixed 
comoving scales in string gas cosmology. The vertical axis is time, 
the horizontal axis is physical distance.  
The solid curve represents the Einstein frame Hubble radius 
$H^{-1}$ which shrinks abruptly to a microphysical scale at $t_R$ and then 
increases linearly in time for $t > t_R$. Fixed comoving scales (the 
dotted lines labeled by $k_1$ and $k_2$) which are currently probed 
in cosmological observations have wavelengths which are smaller than 
the Hubble radius before $t_R$. They exit the Hubble 
radius at times $t_i(k)$ just prior to $t_R$, and propagate with a 
wavelength larger than the Hubble radius until they reenter the 
Hubble radius at times $t_f(k)$.} \label{fig:4} 
\end{figure}

In contrast, in inflationary cosmology (Figure 1) the Hubble radius 
is constant during inflation ($t < t_R$, where here $t_R$ is the 
time of inflationary reheating), whereas the physical wavelength 
corresponding to a fixed comoving scale expands exponentially. Thus, 
as long as the period of inflation is sufficiently long, all scales 
of interest for current cosmological observations are sub-Hubble at 
the beginning of inflation. 
 
There are both important similarities and key differences between
the structure formation mechanisms in inflationary cosmology and string 
gas cosmology. In both cases, scales are sub-Hubble during the early stages,
thus allowing for a causal generation mechanism. Also, in both cases
the fluctuations evolve on super-Hubble scales for a long time in
the radiation phase of standard cosmology, thus leading to their
squeezing, which in turn leads to the phase coherence of the
fluctuations which generate the acoustic oscillations in the angular
power spectrum of the CMB. However, the 
actual generation mechanism for fluctuations is completely different. 
In inflationary cosmology, any thermal fluctuations present before 
the onset of inflation are red-shifted away, leaving us with a quantum 
vacuum state, whereas in the quasi-static Hagedorn phase of string gas 
cosmology matter is in a thermal state. Hence, whereas in inflationary 
cosmology the fluctuations originate as quantum vacuum perturbations, 
in string gas cosmology the inhomogeneities are created by the thermal 
fluctuations of the string gas. 
 
As we have shown in \cite{NBV,Ali,BNPV2}, string thermodynamical
fluctuations in 
the Hagedorn phase of string gas cosmology yield an almost 
scale-invariant spectrum of both scalar and tensor modes. This
result stems from the holographic scaling of the specific heat
$C_V(R)$ (evaluated for fixed volume) as a function of the radius
$R$ of the box
\be \label{specheat}
C_V(R) \, \sim \, R^2 \, .
\ee
As derived in \cite{Deo}, this result holds true for a gas of closed
strings in a space-time in which the three large spatial dimensions
are compact (see \cite{Joao,SA} for recent papers emphasizing the
role of holography). The scaling (\ref{specheat}) is an 
intrinsically stringy result: thermal fluctuations of a gas of 
particles would lead to a very different scaling. 

Since the primordial perturbations in our scenario are of 
thermal origin (and there are no non-vanishing chemical potentials), 
they will be adiabatic. The spectrum of scalar metric fluctuations
has a slight red tilt. As a distinctive feature \cite{BNPV1}, our
scenario predicts a slight blue tilt for the spectrum of gravitational
waves. The red tilt for the scalar modes is due to the fact that the
temperature when short wavelength modes exit the Hubble radius is
slightly lower than the temperature when longer wavelength modes exit.
The gravitational wave amplitude, in contrast, is determined by
the pressure. Since the pressure is closer to zero the deeper in the
Hagedorn phase we are, a slight blue tilt for the tensor fluctuations
results. These results are explained in more detail in a recent review 
\cite{RHBrev7} and in the original references. Here, we will only
focus on some of the key steps. 
   
Our approximation scheme for computing the cosmological 
perturbations and gravitational wave spectra from string gas 
cosmology is as follows (the analysis is similar to how 
the calculations were performed in \cite{BST,BK} in the 
case of inflationary cosmology). For a fixed comoving scale $k$ we follow 
the matter fluctuations until the time $t_i(k)$ 
shortly before the end of the Hagedorn phase when the scale exits 
the Hubble radius \footnote{Recall that on sub-Hubble scales, the
dynamics of matter is the dominant factor in the evolution of the
system, whereas on super-Hubble scales, matter fluctuations freeze
out and gravity dominates. Thus, it is precisely at the
time of Hubble radius crossing that we must extract the
metric fluctuations from the matter perturbations. Since the
concept of an energy density fluctuation is gauge-dependent on
super-Hubble scales, one cannot extrapolate the matter spectra
to larger scales as was done in Section 3 of \cite{KKLM}.} 
At that time, we use the Einstein constraint
equations (discussed below) to compute the values of 
$\Phi(k)$ and $h(k)$ ($h$ is 
the amplitude of the gravitational wave tensor), and then we
propagate the metric fluctuations according to the standard
gravitational perturbation equations until scales re-enter the
Hubble radius at late times.

The first key point is to show how the scalar and tensor metric fluctuations 
can be extracted from knowledge of the energy-momentum tensor of 
the string gas. Working in conformal time $\eta$ and in the longitudinal gauge
for the scalar metric fluctuations,
the metric of a homogeneous and isotropic background 
space-time perturbed by linear cosmological perturbations and 
gravitational waves can be written in the form 
\be \label{pertmetric}
d s^2 = a^2(\eta) \left\{(1 + 2 \Phi)d\eta^2 - [(1 - 
2 \Phi)\delta_{ij} + h_{ij}]d x^i d x^j\right\} \,. 
\ee 
Here, $\Phi$ (which is a function of space and time) describes 
the scalar metric fluctuations \footnote{To avoid confusion with
the dilaton $\phi$, we in this section denote the relativistic
gravitational potential by $\Phi$.}. The tensor $h_{ij}$ is 
transverse and traceless and contains the two polarization states of 
the gravitational waves. We have assumed
that there is no anisotropic stress.

Inserting the metric (\ref{pertmetric}) into the Einstein equations, 
subtracting the background terms and truncating the perturbative expansion 
at linear order leads to the following system of equations 
\begin{eqnarray} \label{perteom3} 
- 3 {\cal H} \left( {\cal H} \Phi + \Phi^{'} \right) + \nabla^2 \Phi 
\, 
&=& \, 4 \pi G a^2 \delta T^0{}_0 \nonumber \\ 
\left( {\cal H} \Phi + \Phi^{'} \right)_{, i} \, 
&=& 4 \pi G a^2 \delta T^0{}_i  \nonumber \\ 
\left[ \left( 2 {\cal H}^{'} + {\cal H}^2 \right) \Phi + 3 {\cal H} 
\Phi^{'} 
+ \Phi^{''} \right] 
&=& - 4 \pi G a^2 \delta T^i{}_i \, , \nonumber \\ 
-{1 \over 2} \left[ {{\cal H}^{\prime}} + {1 \over 
2}{\cal H}^2 \right] h_{ij} 
+ {1 \over 4} {\cal H} h_{ij}^{\prime} && \nonumber \\ 
+ \left[{{\partial^2} \over {\partial \eta^2}} - \nabla^2\right] 
h_{ij} \, 
&=& - 4 \pi G a^2 \delta T^i{}_j \,,\nonumber \\ 
&& \mbox{for $i \neq j$}\, . 
\end{eqnarray} 

In the Hagedorn phase, these equations simplify substantially and 
allow us to extract the scalar and tensor metric fluctuations 
individually. Replacing comoving by 
physical coordinates, we obtain from the $00$ equation 
\be 
\label{scalar} \nabla^2 \Phi \, = \, 4 \pi G \delta T^0{}_0 
\, 
\ee 
and from the $i \neq j$ equation 
\be 
\label{tensor} \nabla^2 h_{ij} \, = \, - 4 \pi G \delta T^i{}_j 
\, . 
\ee 

The above equations (\ref{scalar}) and (\ref{tensor}) allow 
us to compute the power spectra of scalar and tensor metric 
fluctuations in terms of correlation functions of the string
energy-momentum tensor. Since the metric perturbations are 
small in amplitude we can consistently work in Fourier space. 
Specifically, 
\be \label{scalarexp} 
\langle|\Phi(k)|^2\rangle \, = \, 16 \pi^2 G^2 
k^{-4} \langle\delta T^0{}_0(k) \delta T^0{}_0(k)\rangle \, , 
\ee 
where the pointed brackets indicate expectation values, and 
\be 
\label{tensorexp} \langle|h(k)|^2\rangle \, = \, 16 \pi^2 G^2 
k^{-4} \langle\delta T^i{}_j(k) \delta T^i{}_j(k)\rangle \,, 
\ee 
where on the right hand side of (\ref{tensorexp}) we mean the 
average over the correlation functions with $i \neq j$. 

The second key step is to compute the matter fluctuations in
the Hagedorn phase. Since this phase is dominated by the
gas of strings, fluctuations in our scenario are the thermal
fluctuations of a string gas. We will consider a gas of
closed strings in a compact space, i.e. our three-dimensional
space is considered to be large but compact. Specifically, it is
important to have winding modes in the spectrum of string states.

General thermodynamical relations allow us to compute the
matter correlation functions in terms of the the specific heat
$C_V$ and the pressure $p$. The result for the
energy density fluctuation is
\be \label{cor1}
\langle \delta\rho^2 \rangle \, = \,  
- \frac{1}{R^{6}} \frac{\partial}{\partial \beta} 
\left(F + \beta \frac{\partial F}{\partial \beta}\right) \, = \, 
\frac{T^2}{R^6} C_V \, , 
\ee 
where $F$ is the free energy of the string gas and
$\beta$ is the inverse temperature.
The off-diagonal pressure fluctuations, in turn, are given by
\bea \label{cor2} 
\langle \delta {T^i{}_j}^2 
\rangle \, &=& \, \langle {T^i{}_j}^2 \rangle - \langle T^i{}_j \rangle^2 \\ 
&=& \, \frac{1}{\beta R^3}\frac{\partial}{\partial \ln{R}}\left(- 
\frac{1}{R^3} \frac{\partial F}{\partial \ln{R}}\right)  = 
\frac{1}{\beta R^2}\frac{\partial p}{\partial R} \, . \nonumber
\eea 

Now, we apply these relations to the thermodynamics of strings.
In \cite{Deo}, the thermodynamical properties of a gas of closed
strings in a toroidal space of radius $R$ were computed. 
To compute the fluctuations in a region 
of radius $R$ which forms part of our three-dimensional compact
space, we will apply the results of \cite{Deo} to a box of
strings in a volume $V = R^3$. 

The starting point of the computation is the formula for 
the density of states $\Omega(E,R)$ which then determines the 
cpecific heat and the pressure. The result for the specific
heat is
\be \label{specheat2} 
C_V  \, \approx \, 2 \frac{R^2/\ell^3}{T \left(1 - T/T_H\right)}\, . 
\ee 
The `holographic' scaling $C_V(R) \sim R^2$ is responsible for the
overall scale-invariance of the spectrum of cosmological perturbations. 
The factor $(1 - T/T_H)$ in the denominator is responsible 
for giving the spectrum a slight red tilt.

For the pressure. we obtain 
\be 
p(E, R) \approx n_H T_H - \frac{2}{3}\frac{(1 - T/T_H)}{\ell_s^3 
R}\ln{\left[\frac{\ell_s^3 T}{R^2 (1- T/T_H)}\right]} \,, 
\ee 
which immediately yields 
\be \label{tensorresult} 
\langle \delta {T^i{}_j}^2 \rangle \, 
\simeq \, \frac{T (1 - T/T_H)}{\ell_s^3 R^4} 
\ln^2{\left[\frac{R^2}{\ell_s^2}(1 - T/T_H)\right]}\, .  
\ee 
Note that the factor $(1 - T/T_H)$ (which appears in
the numerator of the key term 
in the expression for the microcanonical partition
function) is in the numerator, whereas it was
in the denominator in the expression for the specific heat. The reason
is that in deriving the specific heat from the microcanonical partition
function, a temperature derivative was taken, but not so in deriving
the pressure. This leads to the 
slight blue tilt of the spectrum of gravitational waves, characteristic
of our proposed structure formation scenario. As mentioned
earlier, the physical reason for this blue tilt is that larger wavelength
modes exit the Hubble radius deeper in the Hagedorn phase where the
pressure is smaller and thus the strength of the tensor modes is less.

The third key step in the analysis is to compute the power spectra
for the scalar and tensor modes based on the earlier results.
The power spectrum of scalar metric fluctuations is given by
\bea \label{power2} 
P_{\Phi}(k) \, & \equiv & \, {1 \over {2 \pi^2}} k^3 |\Phi(k)|^2 \\
&=& \, 8 G^2 k^{-1} <|\delta \rho(k)|^2> \, . \nonumber \\
&=& \, 8 G^2 k^2 <(\delta M)^2>_R \nonumber \\ 
               &=& \, 8 G^2 k^{-4} <(\delta \rho)^2>_R 
\, , \nonumber 
\eea 
where in the first step we have used (\ref{scalarexp}) to replace the 
expectation value of $|\Phi(k)|^2$ in terms of the correlation function 
of the energy density, and in the second step we have made the 
transition to position space (note that $k = R^{-1}$).

According to (\ref{cor1}), the density correlation function 
is given by the specific heat via $T^2 R^{-6} C_V$.
Inserting the expression from (\ref{specheat2}) for the specific 
heat of a string gas on a scale $R$ yields to the final result 
\be \label{power4} 
P_{\Phi}(k) \, = \, 8 G^2 {T \over {\ell_s^3}} {1 \over {1 - T/T_H}} 
\ee 
for the power spectrum of cosmological fluctuations. In the above
equation, the temperature $T$ is to be evaluated at the time $t_i(k)$
when the mode $k$ exits the Hubble radius. Since modes with larger
values of $k$ exit the Hubble radius slightly later when the
temperature is slightly lower, a small red tilt of the spectrum is
induced. The amplitude ${\cal A}_S$ of the power spectrum is 
given by
\be
{\cal A}_S \, \sim \, \bigl({{l_{pl}} \over {l_s}}\bigr)^4 
{1 \over {1 - T/T_H}} \, .
\ee
Taking the last factor to be of order unity, we find that a string
length three orders of magnitude larger than the Planck length,
a string length which was assumed in early studies of string theory,
gives the correct amplitude of the spectrum. Thus, it appears that
the string gas cosmology structure formation mechanism does not have
a serious amplitude problem.

Similarly, we can compute the power spectrum of the gravitational waves
and obtain 
\be \label{tpower3} 
P_h(k) \, \sim \, 8 G^2 {T \over {\ell_s^3}} (1 - 
T/T_H) \ln^2{\left[\frac{1}{\ell_s^2 k^2}(1 - T/T_H)^{-1}\right]}\, .  
\ee 
This shows that the spectrum of tensor modes is - to a first approximation, 
namely neglecting the logarithmic factor and neglecting the k-dependence 
of $T(t_i(k))$ - scale-invariant. The k-dependence of the temperature
at Hubble radius crossing induces a small blue tilt for the spectrum
of gravitational waves.

Comparing (\ref{power4}) and (\ref{tpower3}) we see that the tensor
to scalar ratio is suppressed by the factor $(1 - T/T_H)^2$.
Given a good understanding of the exit from the Hagedorn phase we
would be able to compute this ratio as well as the magnitude of the
spectral tilts for both scalar and tensor modes.

\subsection{Discussion}

In order to put our structure formation scenario on a firm basis, we need a
consistent description of the Hagedorn phase. The dilaton gravity
background allows us to understand the onset of the Hagedorn phase
(going backwards in time), but since the dilaton blows up, we rapidly
leave the domain of applicability of the model. 

A background in which our string gas structure formation scenario can
be implemented \cite{Biswas2} is the ghost-free and asymptotically
free higher derivative gravity model proposed in \cite{Biswas} given
by the gravitational action
\be
S \, = \, \int d^4x \sqrt{-g} F(R)
\ee
with
\be
F(R) \, = \, R + \sum_{n = 0}^{\infty} {{c_n} \over {M_s^{2n}}} R 
\bigl( {{\partial^2} \over {\partial^2 t}} - \nabla^2 \bigr)^n R
\, ,
\ee
where $M_s$ is the string mass scale (more generally, it is the scale
where non-perturbative effects start to dominate), and the $c_n$ are
coefficients of order unity.

As shown in \cite{Biswas} and \cite{Biswas2}, this action has bouncing
cosmological solutions. If the temperature during the bounce phase is
sufficiently high, then a gas of strings will be excited in this
phase. In the absence of initial cosmological perturbations in the
contracting phase, our string gas structure formation scenario is
realized. The string network will contain winding modes in the same
way that a string network formed during a cosmological phase transition
will contain infinite strings. The dilaton is fixed in this scenario,
thus putting the calculation of the cosmological perturbations on
a firm basis. There are no additional dynamical degrees of freedom
compared to those in Einstein gravity. The higher derivative
corrections to the equations of motion (in particular to the
Poisson equation) are suppressed by factors of $(k/M_s)^2$. Thus,
all of the conditions on a cosmological backgroound to successfully
realize the string gas cosmology structure formation scenario are
realized.

\section{Conclusions}

These lectures have focused on three topics in theoretical cosmology.
The first is the inflationary universe scenario, the current paradigm
of early universe cosmology. Inflation has been an extremely successful
scenario. It explains why the universe is spatially flat and
isotropic to the extent it is observed to be, and why it is so large
and contains such a large amount of entropy, thus resolving several
mysteries which the previous paradigm of cosmology, the SBB model, was not
able to explain. Possibly more importantly, inflation provides a
causal mechanism for the origin of the fluctuations which are
now mapped out to great accuracy by recent cosmological observations.
Inflation correctly (and fifteen years before the precision observations)
predicted the power spectrum of CMB anisotropies.

The theory of cosmological perturbations is the key tool which allows
us to take theories of the very early universe and calculate predictions
for late time cosmology. This theory is applicable whatever the
paradigm of early universe cosmology might be, and it is the second
topic discussed in these lectures.

In spite of the successes of the inflationary universe scenario, key
conceptual questions remain, making it clear that new input from
fundamental physics is required in order to develop a better theory
of the very early universe. The third topic discussed in these
lectures, string gas cosmology, is an attempt to explore
possible consequences for cosmology of superstring theory, the likely
candidate for the new physics required for early universe cosmology.
In particular, an alternative structure formation scenario not requiring
inflationary dynamics emerges.

\centerline{Acknowledgements}

I wish to thank the organizers of the summer school for inviting
me to lecture and for their hospitality in Dubrovnik. My research
is supported by funds from NSERC and from the Canada Research Chair
program.


\begin{thebibliography}{99}



\bibitem{2dF}
W.~J.~Percival {\it et al.}  [The 2dFGRS Collaboration],   
  ``The 2dF Galaxy Redshift Survey: The power spectrum and the matter content
  of the universe,''
  Mon.\ Not.\ Roy.\ Astron.\ Soc.\  {\bf 327}, 1297 (2001)
  [arXiv:astro-ph/0105252].
  %%CITATION = ASTRO-PH 0105252;%%

\bibitem{SDSS}
C.~Stoughton {\it et al.}  [SDSS Collaboration],
  ``The Sloan Digital Sky Survey: Early data release,''
  Astron.\ J.\  {\bf 123}, 485 (2002).
  %%CITATION = ANJOA,123,485;%%

\bibitem{WMAP}
C.~L.~Bennett {\it et al.},
   ``First Year Wilkinson Microwave Anisotropy Probe (WMAP) Observations:
  Preliminary Maps and Basic Results,''
  Astrophys.\ J.\ Suppl.\  {\bf 148}, 1 (2003)
  [arXiv:astro-ph/0302207].
  %%CITATION = ASTRO-PH 0302207;%%

\bibitem{ShellVil}
A. Vilenkin and E.P.S. Shellard;
\textit{Cosmic Strings and Other Topological Defects},
(Cambridge Univ. Press, Cambridge, 1994).

\bibitem{HK}
M.~B.~Hindmarsh and T.~W.~Kibble,
``Cosmic strings,''
Rept.\ Prog.\ Phys.\  {\bf 58}, 477 (1995)
[arXiv:hep-ph/9411342].
%%CITATION = HEP-PH 9411342;%%

\bibitem{RHBrev0}
R.~H.~Brandenberger,
``Topological defects and structure formation,''
Int.\ J.\ Mod.\ Phys.\ A {\bf 9}, 2117 (1994)
[arXiv:astro-ph/9310041].
%%CITATION = ASTRO-PH 9310041;%%

\bibitem{Cliff}
C.~P.~Burgess,
  ``Inflatable string theory?,''
  Pramana {\bf 63}, 1269 (2004)
  [arXiv:hep-th/0408037].
  %%CITATION = HEP-TH 0408037;%%

\bibitem{Jim}
J.~M.~Cline,
  ``Inflation from string theory,''
  arXiv:hep-th/0501179.
  %%CITATION = HEP-TH 0501179;%%

\bibitem{Andrei}
A.~Linde,
  ``Inflation and string cosmology,''
  eConf {\bf C040802}, L024 (2004)
  [arXiv:hep-th/0503195].
  %%CITATION = HEP-TH 0503195;%%

\bibitem{COBE}
J.~C.~Mather {\it et al.},
   ``A Preliminary Measurement Of The Cosmic Microwave Background Spectrum By
  The Cosmic Background Explorer (Cobe) Satellite,''
  Astrophys.\ J.\  {\bf 354}, L37 (1990).
  %%CITATION = ASJOA,354,L37;%%

\bibitem{Halpern}
H. Gush, M. Halpern and E. Wishnow, 
``Rocket Measurement of the Cosmic-Background-Radiation mm-Wave Spectrum'',
Phys. Rev. Lett. {\bf 65}, 537 (1990).

\bibitem{Guth}
A.~H.~Guth,
  ``The Inflationary Universe: A Possible Solution To The Horizon And Flatness
  Problems,''
  Phys.\ Rev.\ D {\bf 23}, 347 (1981).
  %%CITATION = PHRVA,D23,347;%%

\bibitem{Sato}
K.~Sato,
  ``First Order Phase Transition Of A Vacuum And Expansion Of The Universe,''
  Mon.\ Not.\ Roy.\ Astron.\ Soc.\  {\bf 195}, 467 (1981).
  %%CITATION = MNRAA,195,467;%%

\bibitem{Brout}
R.~Brout, F.~Englert and E.~Gunzig,
  ``The Creation Of The Universe As A Quantum Phenomenon,''
  Annals Phys.\  {\bf 115}, 78 (1978).
  %%CITATION = APNYA,115,78;%%

\bibitem{Starob}
A.~A.~Starobinsky,
  ``A New Type Of Isotropic Cosmological Models Without Singularity,''
  Phys.\ Lett.\ B {\bf 91}, 99 (1980).
  %%CITATION = PHLTA,B91,99;%%

\bibitem{ChibMukh}
V.~F.~Mukhanov and G.~V.~Chibisov,
``Quantum Fluctuation And 'Nonsingular' Universe. (In Russian),''
JETP Lett.\  {\bf 33}, 532 (1981)
[Pisma Zh.\ Eksp.\ Teor.\ Fiz.\  {\bf 33}, 549 (1981)];\\
%%CITATION = JTPLA,33,532;%%

\bibitem{Press}
W. Press,
``Spontaneous production of the Zel'dovich spectrum of cosmological 
fluctuations'',
 Phys. Scr. {\bf 21}, 702 (1980).

\bibitem{kinflation}
C.~Armendariz-Picon, T.~Damour and V.~Mukhanov,
``k-inflation,''
Phys.\ Lett.\ B {\bf 458}, 209 (1999)
[arXiv:hep-th/9904075].
%%CITATION = HEP-TH 9904075;%%

\bibitem{Kung}
R.~H.~Brandenberger and J.~H.~Kung,
``Chaotic Inflation As An Attractor In Initial Condition Space,''
Phys.\ Rev.\ D {\bf 42}, 1008 (1990).
%%CITATION = PHRVA,D42,1008;%%

\bibitem{Goldwirth}
D.~S.~Goldwirth and T.~Piran,
``Initial conditions for inflation,''
Phys.\ Rept.\  {\bf 214}, 223 (1992).
%%CITATION = PRPLC,214,223;%%

\bibitem{Coleman}
S.~R.~Coleman,
``The Fate Of The False Vacuum. 1. Semiclassical Theory,''
Phys.\ Rev.\ D {\bf 15}, 2929 (1977)
[Erratum-ibid.\ D {\bf 16}, 1248 (1977)];\\
%%CITATION = PHRVA,D15,2929;%%
%\cite{Callan:1977pt}
%\bibitem{Callan:1977pt}
C.~G.~.~Callan and S.~R.~Coleman,
``The Fate Of The False Vacuum. 2. First Quantum Corrections,''
Phys.\ Rev.\ D {\bf 16}, 1762 (1977).
%%CITATION = PHRVA,D16,1762;%%

\bibitem{RMP}
R.~H.~Brandenberger,
``Quantum Field Theory Methods And Inflationary Universe Models,''
Rev.\ Mod.\ Phys.\  {\bf 57}, 1 (1985).
%%CITATION = RMPHA,57,1;%%

\bibitem{Kolb}
S.~Dodelson, W.~H.~Kinney and E.~W.~Kolb,
``Cosmic microwave background measurements can discriminate among  inflation
models,''
Phys.\ Rev.\ D {\bf 56}, 3207 (1997)
[arXiv:astro-ph/9702166].
%%CITATION = ASTRO-PH 9702166;%%

\bibitem{new}
A.~D.~Linde,
%``A New Inflationary Universe Scenario: A Possible Solution Of The Horizon,
Flatness, Homogeneity, Isotropy And Primordial Monopole Problems,''
Phys.\ Lett.\ B {\bf 108}, 389 (1982);\\
%%CITATION = PHLTA,B108,389;%%
%\cite{Albrecht:1982wi}
%\bibitem{Albrecht:1982wi}
A.~Albrecht and P.~J.~Steinhardt,
``Cosmology For Grand Unified Theories With Radiatively Induced Symmetry
Breaking,''
Phys.\ Rev.\ Lett.\  {\bf 48}, 1220 (1982).
%%CITATION = PRLTA,48,1220;%%

\bibitem{CW}
S.~R.~Coleman and E.~Weinberg,
``Radiative Corrections As The Origin Of Spontaneous Symmetry Breaking,''
Phys.\ Rev.\ D {\bf 7}, 1888 (1973).
%%CITATION = PHRVA,D7,1888;%%

\bibitem{GhazalScott}
R.~Brandenberger, G.~Geshnizjani and S.~Watson,
``On the initial conditions for brane inflation,''
Phys.\ Rev.\ D {\bf 67}, 123510 (2003)
[arXiv:hep-th/0302222].
%%CITATION = HEP-TH 0302222;%%

\bibitem{chaotic}
%\cite{Linde:1983gd}
%\bibitem{Linde:1983gd}
A.~D.~Linde,
``Chaotic Inflation,''
Phys.\ Lett.\ B {\bf 129}, 177 (1983).
%%CITATION = PHLTA,B129,177;%%

\bibitem{Linderev}
%\cite{Linde:2004kg}
%\bibitem{Linde:2004kg}
A.~Linde,
``Prospects of inflation,''
arXiv:hep-th/0402051.
%%CITATION = HEP-TH 0402051;%%

\bibitem{Feldman}
%\cite{Feldman:1989hh}
%\bibitem{Feldman:1989hh}
H.~A.~Feldman and R.~H.~Brandenberger,
``Chaotic Inflation With Metric And Matter Perturbations,''
Phys.\ Lett.\ B {\bf 227}, 359 (1989);\\
%%CITATION = PHLTA,B227,359;%%
%\cite{Brandenberger:1990xu}
%\bibitem{Brandenberger:1990xu}
R.~H.~Brandenberger, H.~Feldman and J.~Kung,
``Initial Conditions For Chaotic Inflation,''
Phys.\ Scripta {\bf T36}, 64 (1991).
%%CITATION = PHSTB,T36,64;%%

\bibitem{hybrid}
%\cite{Linde:1993cn}
%\bibitem{Linde:1993cn}
A.~D.~Linde,
``Hybrid inflation,''
Phys.\ Rev.\ D {\bf 49}, 748 (1994)
[arXiv:astro-ph/9307002].
%%CITATION = ASTRO-PH 9307002;%%

\bibitem{PolCar}
%\cite{Polchinski:2004ia}
%\bibitem{Polchinski:2004ia}
J.~Polchinski,
``Introduction to cosmic F- and D-strings,''
arXiv:hep-th/0412244.
%%CITATION = HEP-TH 0412244;%%

\bibitem{DolLin}
%\cite{Dolgov:1982th}
%\bibitem{Dolgov:1982th}
A.~D.~Dolgov and A.~D.~Linde,
``Baryon Asymmetry In Inflationary Universe,''
Phys.\ Lett.\ B {\bf 116}, 329 (1982).
%%CITATION = PHLTA,B116,329;%%

\bibitem{AFW}
%\cite{Abbott:1982hn}
%\bibitem{Abbott:1982hn}
L.~F.~Abbott, E.~Farhi and M.~B.~Wise,
``Particle Production In The New Inflationary Cosmology,''
Phys.\ Lett.\ B {\bf 117}, 29 (1982).
%%CITATION = PHLTA,B117,29;%%

\bibitem{TB}
%\cite{Traschen:1990sw}
%\bibitem{Traschen:1990sw}
J.~H.~Traschen and R.~H.~Brandenberger,
``Particle Production During Out-Of-Equilibrium Phase Transitions,''
Phys.\ Rev.\ D {\bf 42}, 2491 (1990).
%%CITATION = PHRVA,D42,2491;%%

\bibitem{KLS}
%\cite{Kofman:1994rk}
%\bibitem{Kofman:1994rk}
L.~Kofman, A.~D.~Linde and A.~A.~Starobinsky,
``Reheating after inflation,''
Phys.\ Rev.\ Lett.\  {\bf 73}, 3195 (1994)
[arXiv:hep-th/9405187].
%%CITATION = HEP-TH 9405187;%%

\bibitem{STB}
%\cite{Shtanov:1994ce}
%\bibitem{Shtanov:1994ce}
Y.~Shtanov, J.~H.~Traschen and R.~H.~Brandenberger,
``Universe reheating after inflation,''
Phys.\ Rev.\ D {\bf 51}, 5438 (1995)
[arXiv:hep-ph/9407247].
%%CITATION = HEP-PH 9407247;%%

\bibitem{KLS2}
%\cite{Kofman:1997yn}
%\bibitem{Kofman:1997yn}
L.~Kofman, A.~D.~Linde and A.~A.~Starobinsky,
``Towards the theory of reheating after inflation,''
Phys.\ Rev.\ D {\bf 56}, 3258 (1997)
[arXiv:hep-ph/9704452].
%%CITATION = HEP-PH 9704452;%%

\bibitem{Lythbook}
A. Liddle and D. Lyth, 
\textit{Cosmological Inflation and Large-Scale Structure},
(Cambridge Univ. Press, Cambridge, 2000).

\bibitem{MFB}
V.~F.~Mukhanov, H.~A.~Feldman and R.~H.~Brandenberger,
  ``Theory Of Cosmological Perturbations. Part 1. Classical Perturbations. Part
  2. Quantum Theory Of Perturbations. Part 3. Extensions,''
  Phys.\ Rept.\  {\bf 215}, 203 (1992).
  %%CITATION = PRPLC,215,203;%%

\bibitem{RHBrev2}
R.~H.~Brandenberger,
  ``Lectures on the theory of cosmological perturbations,''
  Lect.\ Notes Phys.\  {\bf 646}, 127 (2004)
  [arXiv:hep-th/0306071].
  %%CITATION = HEP-TH 0306071;%%

\bibitem{Harrison}
%\cite{Harrison:fb}
%\bibitem{Harrison:fb}
E.~R.~Harrison,
``Fluctuations At The Threshold Of Classical Cosmology,''
Phys.\ Rev.\ D {\bf 1}, 2726 (1970).
%%CITATION = PHRVA,D1,2726;%%

\bibitem{Zeldovich}
%\cite{Zeldovich:1972ij}
%\bibitem{Zeldovich:1972ij}
Y.~B.~Zeldovich,
``A Hypothesis, Unifying The Structure And The Entropy Of The Universe,''
Mon.\ Not.\ Roy.\ Astron.\ Soc.\  {\bf 160}, 1 (1972).
%%CITATION = MNRAA,160,1;%%

\bibitem{Afshordi}
%\cite{Afshordi:2000nr}
%\bibitem{Afshordi:2000nr}
N.~Afshordi and R.~H.~Brandenberger,
``Super-Hubble nonlinear perturbations during inflation,''
Phys.\ Rev.\ D {\bf 63}, 123505 (2001)
[arXiv:gr-qc/0011075].
%%CITATION = GR-QC 0011075;%%

\bibitem{Lifshitz}
%\cite{Lifshitz:1945du}
%\bibitem{Lifshitz:1945du}
E.~Lifshitz,
``On The Gravitational Stability Of The Expanding Universe,''
J.\ Phys.\ (USSR) {\bf 10}, 116 (1946); \\
%%CITATION = JOPYA,10,116;%%
%\cite{Lifshitz:ps}
%\bibitem{Lifshitz:ps}
E.~M.~Lifshitz and I.~M.~Khalatnikov,
``Investigations In Relativistic Cosmology,''
Adv.\ Phys.\  {\bf 12}, 185 (1963).
%%CITATION = ADPHA,12,185;%%

\bibitem{Bardeen}
%\cite{Bardeen:kt}
%\bibitem{Bardeen:kt}
J.~M.~Bardeen,
``Gauge Invariant Cosmological Perturbations,''
Phys.\ Rev.\ D {\bf 22}, 1882 (1980).
%%CITATION = PHRVA,D22,1882;%%

\bibitem{PV}
W. Press and E. Vishniac, 
``Tenacious myths about cosmological perturbations larger than the horizon
  size,''
  Astrophys.\ J.\  {\bf 239}, 1 (1980).
  %%CITATION = ASJOA,239,1;%%

\bibitem{Kodama}
%\cite{Kodama:bj}
%\bibitem{Kodama:bj}
H.~Kodama and M.~Sasaki,
``Cosmological Perturbation Theory,''
Prog.\ Theor.\ Phys.\ Suppl.\  {\bf 78}, 1 (1984).
%%CITATION = PTPSA,78,1;%%

\bibitem{Ellis}
%\cite{Bruni:1991kb}
%\bibitem{Bruni:1991kb}
M.~Bruni, G.~F.~Ellis and P.~K.~Dunsby,
``Gauge invariant perturbations in a scalar field dominated universe,''
Class.\ Quant.\ Grav.\  {\bf 9}, 921 (1992).
%%CITATION = CQGRD,9,921;%%

\bibitem{Hwang}
%\cite{Hwang:1993ai}
%\bibitem{Hwang:1993ai}
J.~c.~Hwang,
``Evolution of ideal fluid cosmological perturbations,''
Astrophys.\ J.\  {\bf 415}, 486 (1993).
%%CITATION = ASJOA,415,486;%% 

\bibitem{Durrer}
R. Durrer,
``Anisotropies in the cosmic microwave background: Theoretical
  foundations,''
  Helv.\ Phys.\ Acta {\bf 69}, 417 (1996)
  [arXiv:astro-ph/9610234].
  %%CITATION = ASTRO-PH 9610234;%%

\bibitem{Stewart}
J. Stewart, 
``Perturbations of Friedmann-Robertson-Walker Cosmological Models'',
Class. Quant. Grav. {\bf 7}, 1169 (1990).

\bibitem{SteWa}
J. Stewart and M. Walker, 
``Perturbations Of Spacetimes In General Relativity,''
  Proc.\ Roy.\ Soc.\ Lond.\ A {\bf 341}, 49 (1974).
  %%CITATION = PRSLA,A341,49;%%

\bibitem{DurSak}
R.~Durrer and M.~Sakellariadou,
  ``A New contribution to cosmological perturbations of some inflationary
  models,''
  Phys.\ Rev.\ D {\bf 50}, 6115 (1994)
  [arXiv:astro-ph/9404043].
  %%CITATION = ASTRO-PH 9404043;%%

\bibitem{BST}
%\cite{Bardeen:qw}
%\bibitem{Bardeen:qw}
J.~M.~Bardeen, P.~J.~Steinhardt and M.~S.~Turner,
``Spontaneous Creation Of Almost Scale - Free Density Perturbations In An Inflationary Universe,''
Phys.\ Rev.\ D {\bf 28}, 679 (1983).
%%CITATION = PHRVA,D28,679;%%

\bibitem{BK}
%\cite{Brandenberger:tg}
%\bibitem{Brandenberger:tg}
R.~H.~Brandenberger and R.~Kahn,
``Cosmological Perturbations In Inflationary Universe Models,''
Phys.\ Rev.\ D {\bf 29}, 2172 (1984).
%%CITATION = PHRVA,D29,2172;%%

\bibitem{Lyth}
%\cite{Lyth:1984gv}
%\bibitem{Lyth:1984gv}
D.~H.~Lyth,
``Large Scale Energy Density Perturbations And Inflation,''
Phys.\ Rev.\ D {\bf 31}, 1792 (1985).
%%CITATION = PHRVA,D31,1792;%%

\bibitem{Fabio1}
F.~Finelli and R.~H.~Brandenberger,
``Parametric amplification of gravitational fluctuations during  reheating,''
Phys.\ Rev.\ Lett.\  {\bf 82}, 1362 (1999)
[arXiv:hep-ph/9809490].
%%CITATION = HEP-PH 9809490;%%

\bibitem{BaVi}
B.~A.~Bassett and F.~Viniegra,
``Massless metric preheating,''
Phys.\ Rev.\ D {\bf 62}, 043507 (2000)
[arXiv:hep-ph/9909353].
%%CITATION = HEP-PH 9909353;%%

\bibitem{Fabio2}
F.~Finelli and R.~H.~Brandenberger,
``Parametric amplification of metric fluctuations during reheating in two  field models,''
Phys.\ Rev.\ D {\bf 62}, 083502 (2000)
[arXiv:hep-ph/0003172].
%%CITATION = HEP-PH 0003172;%%

\bibitem{Weinberg2}
S.~Weinberg,
``Adiabatic modes in cosmology,''
Phys.\ Rev.\ D {\bf 67}, 123504 (2003)
[arXiv:astro-ph/0302326].
%%CITATION = ASTRO-PH 0302326;%%

\bibitem{Zhang}
W.~B.~Lin, X.~H.~Meng and X.~M.~Zhang,
``Adiabatic gravitational perturbation during reheating,''
Phys.\ Rev.\ D {\bf 61}, 121301 (2000)
[arXiv:hep-ph/9912510].
%%CITATION = HEP-PH 9912510;%%

\bibitem{GuthPi}
%\cite{Guth:ec}
%\bibitem{Guth:ec}
A.~H.~Guth and S.~Y.~Pi,
``Fluctuations In The New Inflationary Universe,''
Phys.\ Rev.\ Lett.\  {\bf 49}, 1110 (1982).
%%CITATION = PRLTA,49,1110;%%

\bibitem{Starob4}
%\cite{Starobinsky:ee}
%\bibitem{Starobinsky:ee}
A.~A.~Starobinsky,
``Dynamics Of Phase Transition In The New Inflationary Universe Scenario And Generation Of Perturbations,''
Phys.\ Lett.\ B {\bf 117}, 175 (1982).
%%CITATION = PHLTA,B117,175;%%

\bibitem{Hawking}
%\cite{Hawking:1982cz}
%\bibitem{Hawking:1982cz}
S.~W.~Hawking,
``The Development Of Irregularities In A Single Bubble Inflationary Universe,''
Phys.\ Lett.\ B {\bf 115}, 295 (1982).
%%CITATION = PHLTA,B115,295;%%

\bibitem{Mukh2}
%\cite{Mukhanov:rz}
%\bibitem{Mukhanov:rz}
V.~F.~Mukhanov,
``Gravitational Instability Of The Universe Filled With A Scalar Field,''
JETP Lett.\  {\bf 41}, 493 (1985)
[Pisma Zh.\ Eksp.\ Teor.\ Fiz.\  {\bf 41}, 402 (1985)].
%%CITATION = JTPLA,41,493;%%

\bibitem{Mukh3}
%\cite{Mukhanov:jd}
%\bibitem{Mukhanov:jd}
V.~F.~Mukhanov,
``Quantum Theory Of Gauge Invariant Cosmological Perturbations,''
Sov.\ Phys.\ JETP {\bf 67}, 1297 (1988)
[Zh.\ Eksp.\ Teor.\ Fiz.\  {\bf 94N7}, 1 (1988\ ZETFA,94,1-11.1988)].
%%CITATION = SPHJA,67,1297;%%

\bibitem{Sasaki}
%\cite{Sasaki:1986hm}
%\bibitem{Sasaki:1986hm}
M.~Sasaki,
``Large Scale Quantum Fluctuations In The Inflationary Universe,''
Prog.\ Theor.\ Phys.\  {\bf 76}, 1036 (1986).
%%CITATION = PTPKA,76,1036;%%

\bibitem{Lukash}
%\cite{Lukash:1980iv}
%\bibitem{Lukash:1980iv}
V.~N.~Lukash, Pisma Zh. Eksp. Teor. Fiz. {\bf 31}, 631 (1980);\\
V.~N.~Lukash,
``Production Of Phonons In An Isotropic Universe,''
Sov.\ Phys.\ JETP {\bf 52}, 807 (1980)
[Zh.\ Eksp.\ Teor.\ Fiz.\  {\bf 79},  (1980)].
%%CITATION = SPHJA,52,807;%%

\bibitem{BD} N. Birrell and P.C.W. Davies: \textit{Quantum Fields
in Curved Space}, (Cambridge Univ. Press, Cambridge, 1982).

\bibitem{RB84}
%\cite{Brandenberger:wt}
%\bibitem{Brandenberger:wt}
R.~H.~Brandenberger,
``Quantum Fluctuations As The Source Of Classical Gravitational Perturbations In Inflationary Universe,''
Nucl.\ Phys.\ B {\bf 245}, 328 (1984).
%%CITATION = NUPHA,B245,328;%%

\bibitem{BHill}
%\cite{Brandenberger:fc}
%\bibitem{Brandenberger:fc}
R.~H.~Brandenberger and C.~T.~Hill,
``Energy Density Fluctuations In De Sitter Space,''
Phys.\ Lett.\ B {\bf 179}, 30 (1986).
%%CITATION = PHLTA,B179,30;%%

\bibitem{PolStar}
%\cite{Polarski:1995jg}
%\bibitem{Polarski:1995jg}
D.~Polarski and A.~A.~Starobinsky,
``Semiclassicality and decoherence of cosmological perturbations,''
Class.\ Quant.\ Grav.\  {\bf 13}, 377 (1996)
[arXiv:gr-qc/9504030].
%%CITATION = GR-QC 9504030;%%

\bibitem{SZ}
R.~A.~Sunyaev and Y.~B.~Zeldovich,
  ``Small scale fluctuations of relic radiation,''
  Astrophys.\ Space Sci.\  {\bf 7}, 3 (1970).
  %%CITATION = APSSB,7,3;%%

\bibitem{Peebles}
P.~J.~E.~Peebles and J.~T.~Yu,
  ``Primeval adiabatic perturbation in an expanding universe,''
  Astrophys.\ J.\  {\bf 162}, 815 (1970).
  %%CITATION = ASJOA,162,815;%%

\bibitem{Grishchuk}
%\cite{Grishchuk:1974ny}
%\bibitem{Grishchuk:1974ny}
L.~P.~Grishchuk,
``Amplification Of Gravitational Waves In An Istropic Universe,''
{\it Sov.\ Phys.\ JETP} {\bf 40}, 409 (1975) 
[{\it Zh. Eksp. Teor. Fiz.} {\bf 67}, 825 (1974)].
%%CITATION = SPHJA,40,409;%%

\bibitem{eternal}
A.~D.~Linde,
  ``Eternal Chaotic Inflation,''  
  Mod.\ Phys.\ Lett.\ A {\bf 1}, 81 (1986);\\
  %%CITATION = MPLAE,A1,81;%%
A.~D.~Linde and D.~A.~Linde,
  ``Topological defects as seeds for eternal inflation,''
  Phys.\ Rev.\ D {\bf 50}, 2456 (1994)
  [arXiv:hep-th/9402115].
%%CITATION = HEP-TH 9402115;%%

\bibitem{MSSM}
R.~Allahverdi, K.~Enqvist, J.~Garcia-Bellido and A.~Mazumdar,
  ``Gauge invariant MSSM inflaton,''
  arXiv:hep-ph/0605035.
  %%CITATION = HEP-PH 0605035;%%

\bibitem{Adams}
F.~C.~Adams, K.~Freese and A.~H.~Guth,
  ``Constraints On The Scalar Field Potential In Inflationary Models,''
  Phys.\ Rev.\ D {\bf 43}, 965 (1991).
  %%CITATION = PHRVA,D43,965;%%

\bibitem{CSpapers}
N.~Turok and R.~H.~Brandenberger,
  ``Cosmic Strings And The Formation Of Galaxies And Clusters Of Galaxies,''
  Phys.\ Rev.\ D {\bf 33}, 2175 (1986);\\
%%CITATION = PHRVA,D33,2175;%%
H. Sato, ``Galaxy Formation by Cosmic Strings,''
  Prog. Theor. Phys.\  {\bf 75}, 1342 (1986);\\
A. Stebbins, ``Cosmic Strings and Cold Matter'',
  Ap. J. (Lett.) {\bf 303}, L21 (1986).

\bibitem{CSanis}
A.~Albrecht, D.~Coulson, P.~Ferreira and J.~Magueijo,
  ``Causality and the microwave background,''
  Phys.\ Rev.\ Lett.\  {\bf 76}, 1413 (1996)
  [arXiv:astro-ph/9505030];\\
%%CITATION = ASTRO-PH 9505030;%%
J.~Magueijo, A.~Albrecht, D.~Coulson and P.~Ferreira,
  ``Doppler peaks from active perturbations,''
  Phys.\ Rev.\ Lett.\  {\bf 76}, 2617 (1996)
  [arXiv:astro-ph/9511042];\\
%%CITATION = ASTRO-PH 9511042;%%
U.~L.~Pen, U.~Seljak and N.~Turok,
  ``Power spectra in global defect theories of cosmic structure formation,''
  Phys.\ Rev.\ Lett.\  {\bf 79}, 1611 (1997)
  [arXiv:astro-ph/9704165].
%%CITATION = ASTRO-PH 9704165;%%

\bibitem{RHBrev1}
R.~H.~Brandenberger,  
``Inflationary cosmology: Progress and problems,'' publ. in proc. of
IPM School On Cosmology 1999: Large Scale Structure Formation,
  arXiv:hep-ph/9910410.  
%%CITATION = HEP-PH 9910410;%%

\bibitem{Jerome1}
R.~H.~Brandenberger and J.~Martin, ``The robustness of inflation to changes in super-Planck-scale physics,''
Mod.~Phys.~Lett.~A~{\bf 16}, 999 (2001), 
[arXiv:astro-ph/0005432];\\
%%CITATION = ASTRO-PH 0005432;%% 
J.~Martin and R.~H.~Brandenberger,
``The trans-Planckian problem of inflationary cosmology,''
Phys.~Rev.~D~{\bf 63}, 123501 (2001), 
[arXiv:hep-th/0005209].
%%CITATION = HEP-TH 0005209;%%

\bibitem{Niemeyer}
J.~C.~Niemeyer,
``Inflation with a high frequency cutoff,''
Phys.~Rev.~D~{\bf 63}, 123502 (2001),
[arXiv:astro-ph/0005533]; \\
%%CITATION = ASTRO-PH 0005533;%%
S.~Shankaranarayanan, 
``Is there an imprint of Planck scale physics on inflationary cosmology?,''
Class.~Quant.~Grav.~{\bf 20}, 75 (2003), 
[arXiv:gr-qc/0203060];\\
%%CITATION = GR-QC 0203060;%%
J.~C.~Niemeyer and R.~Parentani,
  ``Trans-Planckian dispersion and scale-invariance of inflationary
  perturbations,''
 Phys.~Rev.~D~{\bf 64}, 101301 (2001),
[arXiv:astro-ph/0101451].
%%CITATION = ASTRO-PH 0101451;%%

\bibitem{Unruh}
W.~G.~Unruh,
  ``Sonic analog of black holes and the effects of high frequencies on black
  hole evaporation,''
  Phys.\ Rev.\ D {\bf 51}, 2827 (1995).
  %%CITATION = PHRVA,D51,2827;%%

\bibitem{CJ}
S.~Corley and T.~Jacobson,
  ``Hawking Spectrum and High Frequency Dispersion,''
  Phys.\ Rev.\ D {\bf 54}, 1568 (1996)
  [arXiv:hep-th/9601073].
  %%CITATION = HEP-TH 9601073;%%

\bibitem{Jerome3}
R.~H.~Brandenberger and J.~Martin,
  ``Back-reaction and the trans-Planckian problem of inflation revisited,''
  Phys.\ Rev.\ D {\bf 71}, 023504 (2005)
  [arXiv:hep-th/0410223].
  %%CITATION = HEP-TH 0410223;%%

\bibitem{Tanaka}
T.~Tanaka, 
``A comment on trans-Planckian physics in inflationary universe,''
[arXiv:astro-ph/0012431].
 %%CITATION = ASTRO-PH 0012431;%%

\bibitem{Starob3}
A.~A.~Starobinsky, 
``Robustness of the inflationary perturbation spectrum to trans-Planckian   
physics,''
Pisma Zh.~Eksp.~Teor.~Fiz.~{\bf 73}, 415 (2001),
[JETP Lett.\ {\bf 73}, 371 (2001)], [arXiv:astro-ph/0104043].
%%CITATION = ASTRO-PH 0104043;%%

\bibitem{HE}
S. Hawking and G. Ellis, \textit{The Large-Scale Structure of Space-Time}
(Cambridge Univ. Press, Cambridge, 1973).

\bibitem{Borde}
A.~Borde and A.~Vilenkin,
  ``Eternal inflation and the initial singularity,''
  Phys.\ Rev.\ Lett.\  {\bf 72}, 3305 (1994)
  [arXiv:gr-qc/9312022].
  %%CITATION = GR-QC 9312022;%%

\bibitem{swamp} N.~Arkani-Hamed, L.~Motl, A.~Nicolis and C.~Vafa,
  ``The string landscape, black holes and gravity as the weakest force,''
  arXiv:hep-th/0601001.
  %%CITATION = HEP-TH 0601001;%%

\bibitem{BMS}
R.~H.~Brandenberger, V.~F.~Mukhanov and A.~Sornborger,
  ``A Cosmological theory without singularities,''
  Phys.\ Rev.\ D {\bf 48}, 1629 (1993)
  [arXiv:gr-qc/9303001];\\
  %%CITATION = GR-QC 9303001;%%
V.~F.~Mukhanov and R.~H.~Brandenberger,
  ``A Nonsingular universe,''
  Phys.\ Rev.\ Lett.\  {\bf 68}, 1969 (1992).
  %%CITATION = PRLTA,68,1969;%%

\bibitem{Biswas}
T.~Biswas, A.~Mazumdar and W.~Siegel,
  ``Bouncing universes in string-inspired gravity,''
  JCAP {\bf 0603}, 009 (2006)
  [arXiv:hep-th/0508194].
  %%CITATION = HEP-TH 0508194;%%item{Bojowald}

\bibitem{Bojowald}
M.~Bojowald,   
``Loop quantum cosmology,''   
Living Rev.\ Rel.\  {\bf 8}, 11 (2005)   
[arXiv:gr-qc/0601085].   
%%CITATION = GR-QC 0601085;%%

\bibitem{Hagedorn}
R.~Hagedorn,
  ``Statistical Thermodynamics Of Strong Interactions At High-Energies,''
  Nuovo Cim.\ Suppl.\  {\bf 3}, 147 (1965).
  %%CITATION = NUCUA,3,147;%%

\bibitem{BV}
R.~H.~Brandenberger and C.~Vafa,
  ``Superstrings In The Early Universe,''
  Nucl.\ Phys.\ B {\bf 316}, 391 (1989).
  %%CITATION = NUPHA,B316,391;%%

\bibitem{RHBrev4}
R.~H.~Brandenberger,
``Back reaction of cosmological perturbations and the cosmological constant
problem,'' publ. in the proc. of the
18th IAP Colloquium On The Nature Of Dark Energy: Observational And Theoretical Results On The Accelerating Universe,
arXiv:hep-th/0210165.
%%CITATION = HEP-TH 0210165;%%

\bibitem{branereview}
P.~Brax, C.~van de Bruck and A.~C.~Davis,
  ``Brane world cosmology,''
  Rept.\ Prog.\ Phys.\  {\bf 67}, 2183 (2004)
  [arXiv:hep-th/0404011].
  %%CITATION = HEP-TH 0404011;%%

\bibitem{Mahbub}
M.~Majumdar and A.-C. Davis,
  ``Cosmological creation of D-branes and anti-D-branes,''
  JHEP {\bf 0203}, 056 (2002)
  [arXiv:hep-th/0202148].
  %%CITATION = HEP-TH 0202148;%%

\bibitem{Mairi}
R.~Durrer, M.~Kunz and M.~Sakellariadou,
  ``Why do we live in 3+1 dimensions?,''
  Phys.\ Lett.\ B {\bf 614}, 125 (2005)
  [arXiv:hep-th/0501163].
  %%CITATION = HEP-TH 0501163;%%

\bibitem{Lisa}
A.~Karch and L.~Randall,
  ``Relaxing to three dimensions,''
  Phys.\ Rev.\ Lett.\  {\bf 95}, 161601 (2005)
  [arXiv:hep-th/0506053].
  %%CITATION = HEP-TH 0506053;%%

\bibitem{TV}
A.~A.~Tseytlin and C.~Vafa,
  ``Elements of string cosmology,''
  Nucl.\ Phys.\ B {\bf 372}, 443 (1992)
  [arXiv:hep-th/9109048].
  %%CITATION = HEP-TH 9109048;%%

\bibitem{ABE}
S.~Alexander, R.~H.~Brandenberger and D.~Easson,
  ``Brane gases in the early universe,''
  Phys.\ Rev.\ D {\bf 62}, 103509 (2000)
  [arXiv:hep-th/0005212].
  %%CITATION = HEP-TH 0005212;%%

\bibitem{Rama}
S.~Kalyana Rama,
  ``A principle to determine the number (3+1) of large spacetime dimensions,''
  arXiv:hep-th/0610071.
  %%CITATION = HEP-TH 0610071;%%
  
\bibitem{Perlt}
J.~Kripfganz and H.~Perlt,
  ``Cosmological Impact Of Winding Strings,''
  Class.\ Quant.\ Grav.\  {\bf 5}, 453 (1988).
  %%CITATION = CQGRD,5,453;%%

\bibitem{BattWat}
T.~Battefeld and S.~Watson,   
``String gas cosmology,''   
Rev.\ Mod.\ Phys.\  {\bf 78}, 435 (2006)   
[arXiv:hep-th/0510022].   
%%CITATION = HEP-TH 0510022;%%

\bibitem{RHBrev5}
R.~H.~Brandenberger,   
``Challenges for string gas cosmology,'' publ. in proc. of the   
59th Yamada Conference On Inflating Horizon Of Particle Astrophysics And Cosmology, 
arXiv:hep-th/0509099.   
%%CITATION = HEP-TH 0509099;%%

\bibitem{RHBrev6}
R.~H.~Brandenberger,   
``Moduli stabilization in string gas cosmology,''   
Prog.\ Theor.\ Phys.\ Suppl.\  {\bf 163}, 358 (2006)   
[arXiv:hep-th/0509159].   
%%CITATION = HEP-TH 0509159;%%

\bibitem{RHBrev7}
R.~H.~Brandenberger,
``Conceptual Problems of Inflationary Cosmology and a New Approach to
Cosmological Structure Formation,''
arXiv:hep-th/0701111,
to be publ. in the proceedings of {\it Inflation + 25} (Springer,
Berlin, 2007).
%%CITATION = HEP-TH 0701111;%%

\bibitem{Pol}
J. Polchinski, \textit{String Theory, Vols. 1 and 2},
(Cambridge Univ. Press, Cambridge, 1998).

\bibitem{Boehm}
T.~Boehm and R.~Brandenberger,
  ``On T-duality in brane gas cosmology,''
  JCAP {\bf 0306}, 008 (2003)
  [arXiv:hep-th/0208188].
  %%CITATION = HEP-TH 0208188;%%

\bibitem{Biswas2}
T.~Biswas, R.~Brandenberger, A.~Mazumdar and W.~Siegel,   
``Non-perturbative gravity, Hagedorn bounce and CMB,''   
arXiv:hep-th/0610274.   
%%CITATION = HEP-TH 0610274;%%

\bibitem{gellis}
G.~F.~R.~Ellis and R.~Maartens,   
  ``The emergent universe: Inflationary cosmology with no singularity,''      
  Class.\ Quant.\ Grav.\  {\bf 21}, 223 (2004)   
  [arXiv:gr-qc/0211082];\\  
  %%CITATION = GR-QC 0211082;%%
G.~F.~R.~Ellis, J.~Murugan and C.~G.~Tsagas,
  ``The emergent universe: An explicit construction,''
  Class.\ Quant.\ Grav.\  {\bf 21}, 233 (2004)
  [arXiv:gr-qc/0307112].
  %%CITATION = GR-QC 0307112;%%
  
\bibitem{Kabat3}
R.~Easther, B.~R.~Greene, M.~G.~Jackson and D.~Kabat,
  ``String windings in the early universe,''
  JCAP {\bf 0502}, 009 (2005)
  [arXiv:hep-th/0409121].
  %%CITATION = HEP-TH 0409121;%%

\bibitem{Danos}
R.~Danos, A.~R.~Frey and A.~Mazumdar,
  ``Interaction rates in string gas cosmology,''
  Phys.\ Rev.\ D {\bf 70}, 106010 (2004)
  [arXiv:hep-th/0409162].
  %%CITATION = HEP-TH 0409162;%%

\bibitem{Watson1}
S.~Watson and R.~H.~Brandenberger,
  ``Isotropization in brane gas cosmology,''
  Phys.\ Rev.\ D {\bf 67}, 043510 (2003)
  [arXiv:hep-th/0207168].
  %%CITATION = HEP-TH 0207168;%%

\bibitem{Ven}
G.~Veneziano,
  ``Scale factor duality for classical and quantum strings,''
  Phys.\ Lett.\ B {\bf 265}, 287 (1991).
  %%CITATION = PHLTA,B265,287;%%

\bibitem{Riotto}
M.~Maggiore and A.~Riotto,
  ``D-branes and cosmology,''
  Nucl.\ Phys.\ B {\bf 548}, 427 (1999)
  [arXiv:hep-th/9811089].
  %%CITATION = HEP-TH 9811089;%%

\bibitem{Betal}
R.~H.~Brandenberger {\it et al.},
  ``More on the spectrum of perturbations in string gas cosmology,''
  JCAP {\bf 0611}, 009 (2006)
  [arXiv:hep-th/0608186].
  %%CITATION = HEP-TH 0608186;%%

\bibitem{Kaya06}
S.~Arapoglu, A.~Karakci and A.~Kaya,
  ``S-duality in string gas cosmology,''
  arXiv:hep-th/0611193.
  %%CITATION = HEP-TH 0611193;%%

\bibitem{BEK}
R.~Brandenberger, D.~A.~Easson and D.~Kimberly,
  ``Loitering phase in brane gas cosmology,''
  Nucl.\ Phys.\ B {\bf 623}, 421 (2002)
  [arXiv:hep-th/0109165].
  %%CITATION = HEP-TH 0109165;%%

\bibitem{Natalia}
R.~Brandenberger and N.~Shuhmaher,
  ``The Confining Heterotic Brane Gas: A Non-Inflationary Solution to the
  Entropy and Horizon Problems of Standard Cosmology,''
  JHEP {\bf 0601}, 074 (2006)
  [arXiv:hep-th/0511299].
  %%CITATION = HEP-TH 0511299;%%

\bibitem{NBV}
A.~Nayeri, R.~H.~Brandenberger and C.~Vafa,
  ``Producing a scale-invariant spectrum of perturbations in a Hagedorn phase
  of string cosmology,''
  Phys.\ Rev.\ Lett.\  {\bf 97}, 021302 (2006)
  [arXiv:hep-th/0511140].
  %%CITATION = HEP-TH 0511140;%%

\bibitem{Ali}
A.~Nayeri,
   ``Inflation free, stringy generation of scale-invariant cosmological
  fluctuations in D = 3 + 1 dimensions,''
  arXiv:hep-th/0607073.
  %%CITATION = HEP-TH 0607073;%%

\bibitem{BNPV2}
R. H. Brandenberger, A. Nayeri, S. P. Patil and C. Vafa,
``String gas cosmology and structure formation,''
arXiv:hep-th/0608121.
%%CITATION = HEP-TH 0608121;%%

\bibitem{Deo}
 N.~Deo, S.~Jain, O.~Narayan and C.~I.~Tan,
  ``The Effect of topology on the thermodynamic limit for a string gas,''
  Phys.\ Rev.\ D {\bf 45}, 3641 (1992).
  %%CITATION = PHRVA,D45,3641;%%

\bibitem{BNPV1}
R.~H.~Brandenberger, A.~Nayeri, S.~P.~Patil and C.~Vafa,
  ``Tensor modes from a primordial Hagedorn phase of string cosmology,''
  arXiv:hep-th/0604126.
  %%CITATION = HEP-TH 0604126;%%

\bibitem{KKLM}
N.~Kaloper, L.~Kofman, A.~Linde and V.~Mukhanov,   
``On the new string theory inspired mechanism of generation of cosmological   
perturbations,''   
JCAP {\bf 0610}, 006 (2006)   
[arXiv:hep-th/0608200].   
%%CITATION = HEP-TH 0608200;%%

\bibitem{Joao} 
J.~Magueijo, L.~Smolin and C.~R.~Contaldi,
  ``Holography and the scale-invariance of density fluctuations,''
  arXiv:astro-ph/0611695.
  %%CITATION = ASTRO-PH 0611695;%%

\bibitem{SA} S. Patil and A. Mazumdar, in preparation (2007).

\end{thebibliography}
\end{document}